\documentclass[10pt, final, journal, letterpaper, twocolumn]{IEEEtran}

\makeatletter
\def\ps@headings{%
\def\@oddhead{\mbox{}\scriptsize\rightmark \hfil \thepage}%
\def\@evenhead{\scriptsize\thepage \hfil \leftmark\mbox{}}%
\def\@oddfoot{}%
\def\@evenfoot{}}
\makeatother 

\IEEEoverridecommandlockouts
\usepackage{mdframed}
\usepackage{multirow}
\usepackage{epstopdf}
\usepackage{url}
\usepackage{subfigure}
\usepackage{amsfonts}
\usepackage[dvips]{graphicx}
\usepackage{times}
\usepackage{cite}
\usepackage{amsmath}
\usepackage{array}
\usepackage{amssymb}

\newcommand{\bs}{\boldsymbol}

\usepackage{stfloats}
\usepackage{graphicx}
\usepackage{footnote}
\usepackage{mdframed}
\usepackage{booktabs}
\usepackage{amsthm}
\usepackage{array}
\usepackage{algorithmic}
\usepackage{subeqnarray}
\usepackage{cases}
\usepackage{threeparttable}
\usepackage{color}
\makeatletter
\newif\if@restonecol
\makeatother

\usepackage[linesnumbered,lined,vlined,ruled,commentsnumbered]{algorithm2e}
\usepackage{mdframed}
\newtheorem{theorem}{Theorem}
\newtheorem{remark}{Remark}
\newtheorem{proposition}{Proposition}
\newtheorem{lemma}{Lemma}
\newtheorem*{game}{DeNUM Game}
\newtheorem{definition}{Definition}
\newtheorem{assumption}{Assumption}

\newtheorem{example}{Example}
\newtheorem*{question}{Question}
\newtheorem{mechanism}{Mechanism}
\def\qed{\rule{1\linewidth}{0.75pt}}
\ifodd 0
\newcommand{\rev}[1]{{\color{blue}#1}}

\newcommand{\com}[1]{\textbf{\color{red} (Comment: #1) }}
\newcommand{\comg}[1]{\textbf{\color{blue} (COMMENT: #1)}}
\newcommand{\response}[1]{\textbf{\color{blue} (RESPONSE: #1)}}
\else
\newcommand{\rev}[1]{#1}

\newcommand{\com}[1]{}
\newcommand{\comg}[1]{}
\newcommand{\response}[1]{}
\fi

\begin{document}

\title{Efficient Network Sharing with Asymmetric Constraint Information \vspace{-0.05cm}}
\author{
Meng Zhang and
Jianwei Huang\\

\IEEEcompsocitemizethanks{
This work is supported by the General Research Fund CUHK 14219016 from Hong Kong UGC, the Presidential Fund from the Chinese University of Hong Kong, Shenzhen, and the Shenzhen Institute of Artificial Intelligence and Robotics for Society (AIRS). This paper was presented in part at the IEEE INFOCOM, Paris, France, May 2019 \cite{INFOCOM}. \textit{(Corresponding author: Jianwei Huang.)}

M. Zhang and J. Huang are with Department of Information Engineering, The Chinese University of Hong Kong, Shatin, NT, Hong Kong,  China.  J. Huang is also with School of Engineering and Science, The Chinese University of Hong Kong, Shenzhen, China, and the Shenzhen Institute of Artificial Intelligence and Robotics for Society (AIRS). E-mail: \{zm015, jwhuang\}@ie.cuhk.edu.hk.
}

\vspace{-0.7cm}

}

\maketitle

\thispagestyle{empty}
\begin{abstract}	
Network sharing has become a key feature of various enablers of the next generation network, such as network function virtualization and fog computing architectures. Network utility maximization (NUM) is a general framework for achieving fair, efficient, and cost-effective sharing of constrained network resources.  When agents have asymmetric and private information, however, a fundamental economic challenge is how to solve the NUM Problem considering the self-interests of strategic agents. Many previous related works  have proposed economic mechanisms that can cope with agents' private utilities. However, the network sharing paradigm introduces the issue of information asymmetries regarding constraints. The related literature largely neglected such an issue; limited closely related studies provided solutions only applicable to specific application scenarios. To tackle these issues,  we propose  the Decomposable NUM (DeNUM) Mechanism and the Dynamic DeNUM (DyDeNUM) Mechanism, the first mechanisms in the literature for solving NUM Problems considering private utility and constraint information. The key idea of both mechanisms is to decentralize the decision process to agents, who will make resource allocation decisions  without the need of revealing private information to others. Under a monitorable influence assumption, the DeNUM Mechanism yields the network-utility maximizing solution at an equilibrium, and achieves other desirable economic properties (such as  individual rationality and budget balance). We further establish the connection between the equilibrium structure and the primal-dual solution to a related optimization problem, based on which we prove the convergence of the DeNUM Algorithm to an equilibrium. When the agents' influences are not monitorable, we propose the DyDeNUM Mechanism that yields the network-utility maximizing solution at the cost of the balanced budget. Finally, as a case study, we apply the proposed mechanisms to solving the NUM problem for a fog-based user-provided network, and show that both mechanisms  improve the network utility by $34\%$ compared to a non-cooperation benchmark.
\end{abstract}
\vspace{0.1cm}

\section{Introduction}
\subsection{Motivations}

The proliferation of mobile devices and applications has been significantly increasing the demand for wireless services. According to Cisco, global mobile traffic has been predicted to increase with an annual growth rate of $60\%$ in the next several years, reaching  $48$ exabytes per month in 2021 \cite{Cisco}. 
	The unprecedented traffic demand has been pushing mobile network operators to explore more cost-effective and efficient approaches to provide mobile services.
\emph{Network sharing} is a promising paradigm to reduce capital expenditure and the operational expenditure and achieve efficient network sources utilization. 
It has
 emerged as  an indispensable feature in the 5G system and its enabling architectures including network slicing \cite{NetSlice}, network function virtualization\cite{NFV}, and fog-based networking \cite{Fog}. 

To achieve efficient network sharing, network utility maximization (NUM) is a promising general framework for sharing multiple divisible resources (i.e., those that can be infinitely divided, e.g.,
 bandwidth, power, storages, and network slices) among multiple agents (such as tenants in network slicing architecture and fog nodes in the fog networking architecture) in many network resource allocation problems \cite{NUMTut,layering}. Typically, a NUM Problem aims to optimize allocative/sharing decisions to maximize the aggregate agents' utility, subject to some (coupling) system-level and (uncoupling) local constraints. 
It had found numerous applications across many different areas besides the network sharing applications.\footnote{Examples include wireless sensor networks \cite{Sensor}, mobile networks \cite{comm,Open}, power grids \cite{smartgrid}, and cloud computing networks \cite{Cloud}.}


\setcounter{page}{1}

In practice, a system designer (such as a 5G network slice broker \cite{NetSlice} in the network slicing architecture) of a networked system  does not have complete network information  to solve the NUM directly. Even if agents are willing to share their information, gathering such information by a centralized decision maker can incur significant communication overhands and solving such a problem can lead to significant computational overhead, when the size of the NUM Problem is large. Fortunately, many NUM Problems exhibit the \textit{decomposability structure}  (to be explained in details in Section \ref{Sysm}), which makes it possible to decompose the original centralized NUM Problem into several subproblems \cite{NUMTut}. 
With such a structure, one can design a distributed optimization algorithm through distributively solving subproblems coordinated by proper signaling  (often coinciding with the dual variables \cite{NUMTut}).
Therefore, such a distributed optimization approach can significantly relieve the system designer's burdens of computation and communications.


The distributed optimization approach assumes that agents are obedient, i.e., willing to follow the algorithm. 
However, in practice, an agent can be strategic and self-interested (having her own local objective that is different from the system level objective). Thus, an agent  may attempt to misreport information or tamper with the algorithms to her advantage, which may result in severe allocation inefficiency.
One way for the system designer to address this issue is to design a proper economic mechanism by anticipating such strategic behaviors. For the networked divisible resource allocation problems, related research efforts have mainly focused on the Nash mechanisms which achieve the efficient allocations in a Nash equilibrium  (NE)  (e.g. \cite{VCGKelly,scalar,Jain2010,Multicast,power,Surrogate,learning,PEV,sharma2012local,General}). 


Nevertheless, the network sharing paradigm has introduced several important issues that have been overlooked in
the existing mechanisms in the literature.
First, although most existing mechanisms  (e.g. \cite{VCGKelly,scalar,Jain2010,Multicast,power,Surrogate,learning,PEV,sharma2012local,General}) can cope with strategic agents' private utilities, they assumed that the information regarding the system and local constraints (such as the network topology and capacities) are known by the designer of the mechanism. This is not always true in the network sharing paradigm, since the system designer often does not own the network resources by itself and hence
	 has limited information about the networks. 
	Each self-interested agent may also misreport her private information related to  constraints 
 to her advantage. Misreporting constraint information can also incur severe inefficiency loss, as demonstrated in Section \ref{Illu}.


Second, existing mechanisms proposed for network resource allocation are often  applicable to only specific networking scenarios (e.g. flow control problems \cite{VCGKelly,scalar,Jain2010,Multicast,Surrogate}, power and spectrum allocation \cite{power}, and electric vehicles systems \cite{PEV}). 
These mechanisms often do not work for more general and sophisticated NUM Problems or the general network sharing framework.

%

The above issues motivate the following key question  in this paper: 
\begin{question}
How should one
design a unified mechanism framework for  the NUM problem, considering strategic agents' private information (of both utilities and constraints)?
\end{question}



\subsection{Solution Approach and Contributions}

In this paper, we adopt the idea of optimization decomposition \cite{NUMTut} in the mechanism design, building upon which we first propose a Nash Mechanism for the class of \underline{De}composable \underline{NUM}  (DeNUM) Problems, and we call it the DeNUM mechanism. 
Our approach differs from  the traditional mechanism design approach  in the following sense. 	A traditional mechanism directly determines the allocation and money transfer based on agents' submitted messages \cite{Implementation}. 
In contrast, by exploiting an \textit{indirect optimization decomposition} structure, our DeNUM mechanism decentralizes the allocative decisions to the side of agents. Specifically,  based on agents' submitted messages, the DeNUM Mechanism partitions the system constraints into several 
individual constraints
which are imposed to corresponding agents. Then, the mechanism let agents distributively determine the allocations. Such decentralization eliminates the necessity for agents to reveal their utility and constraint information.
Furthermore, such a constraint partitioning works for any decomposable NUM Problem, and thus constitutes a general mechanism framework.

The success of a Nash mechanism  (such as our proposed DeNUM Mechanism) relies on a distributed algorithm for agents to attain an equilibrium. Imposing individual constraints induces the generalized Nash equilibrium (GNE)  concept, in which agents have interdependent strategy spaces \cite{GNEP}. Designing a distributed algorithm that converges to a GNE is notoriously difficult, since some commonly used NE seeking algorithms fail to converge here \cite{GNEP}.
 We overcome this challenge by establishing the connection between the GNE and the primal-dual solution to a related optimization problem, which makes it possible to design a family of algorithms that can converge to the GNE.

\begin{table*}[]
	\caption{Mechanism Design for the NUM Problems and Related Network Applications}\label{liter}
	\vspace{-5pt}\hspace{0.5cm}
	\scriptsize{\begin{tabular}{|c|c|c|c|c|c|}
			\hline
			\multirow{2}{*}{Reference}                  & \multirow{2}{*}{Framework Type}              & \multirow{2}{*}{Private Constraints} & \multicolumn{2}{c|}{Property}        & \multirow{2}{*}{Distributed Algorithm} \\ \cline{4-5}
			&                                              &                                                 & Full Implementation & Budget Balance &                                   \\ \hline
			\hline
			\multicolumn{6}{|c|}{Nash Mechanisms} \\
			
			\hline
			\cite{VCGKelly,scalar,Jain2010} & Flow Control                                 & $\times$                                        & $\times$            & $\times$       & $\times$                          \\ 
			\cite{Multicast}                            & Flow Control                                 & $\times$                                        & \checkmark          & \checkmark     & $\times$                          \\ 
			\cite{Surrogate}                            & Joint Flow Control and Multi-Path Routing    & $\times$                                        & \checkmark          & \checkmark     & \checkmark                        \\ 
			\cite{power}                                & Power Allocation and Spectrum Sharing        & $\times$                                        & $\times$            & \checkmark     & $\times$                          \\ 
			\cite{PEV}                                  & Electricity Management for Electric Vehicles & $\times$                                        & \checkmark          & $\times$       & $\times$                          \\ 
			\cite{sharma2012local}                      & Networked Public Goods                       & Only Local Constraints                          & \checkmark          & \checkmark     & $\times$                          \\ 
			\cite{learning}                             & Networked Private Goods                      & $\times$                                        & \checkmark          & \checkmark     & \checkmark                        \\ 
			\cite{General}                              & NUM Problems with Linear Constraints         & $\times$                                        & \checkmark          & \checkmark     & $\times$                          \\ \hline
			\textbf{DeNUM}                                  & \textbf{Decomposable NUM Problems}                    & \textbf{\checkmark}                                      & \textbf{\checkmark}          & \textbf{\checkmark}     & \textbf{\checkmark}                       \\ \hline 
			\hline
			\multicolumn{6}{|c|}{Dynamic Mechanisms} \\
			\hline
			\cite{Dynamic1}                           & Flow Control         & $\times$                                        & $\times$             & $\times$       &\textbf{\checkmark}                            \\ 
			\hline
			\cite{Dynamic2}                           & Rate Allocation    & $\times$                                        & $\times$             & $\times$       & \textbf{\checkmark}                            \\ \hline
			\textbf{DyDeNUM}                                  & \textbf{Decomposable NUM Problems}                    & \textbf{\checkmark}                                      & $\times$             & $\times$       & \textbf{\checkmark}                       \\ \hline
		\end{tabular}}
		\vspace{-5pt}
	\end{table*}

Our proposed DeNUM Mechanism assumes that the system designer or the other agents can monitor the influences of each agent's action to the system (such as consuming resources or generating interference). However, in some applications, monitoring might be too costly or difficult. This further motivates us to propose a Dynamic DeNUM (DyDeNUM) Mechanism. Different from the DeNUM Mechanism, the DyDeNUM Mechanism exploits a \textit{direct optimization decomposition} structure that does not further introduce auxiliary constraints. This eliminates the necessity of the monitorable influences. We then show that
 the DyDeNUM Mechanism
 can yield the network utility maximization at an equilibrium
	even when the influence functions are not monitorable. However, such a property comes at a cost of the budget balance.

To summarize, our main contributions are:
\begin{itemize}
	\item \textit{General network sharing mechanism framework}:  Our DeNUM framework, including both the DeNUM Mechanism and the DyDeNUM Mechanism,  together with the related distributed algorithms,  achieves the network utility maximization
	for a general class of NUM Problems.
	\item \textit{Private constraint information:} To the best of our knowledge, we design the first mechanisms in the literature that can cope with agents' information asymmetries regarding system and local constraints in additional to the asymmetric utility information. 
	\item \textit{Distributed algorithm design:} For agents to distributively attach the GNE of the DeNUM Mechanism, we
	further propose the DeNUM Algorithm. We prove its convergence by relating the GNE to the primal-dual solution to a related optimization problem. Such a proof methodology also  suggests a general approach to designing distributed algorithms.
	\item \textit{Elimination of monitorability requirement:} \rev{Our DyDeNUM Mechanism can achieve the network utility maximizing outcome at an equilibrium even if agents' influences are not monitorable, at the cost of the budget balance.}
	\item \textit{Fog-based user-provided network:} We apply the DeNUM framework  to the fog-based application  user-provided networks, of  which existing mechanisms are inapplicable. We show that both mechanisms can improve the network utility by $34\%$ compared to a benchmark.	
\end{itemize}

 We organize the rest of this paper as follows. We review the literature in Section \ref{related}, and motivates our study in Section \ref{Illu} with an example of system inefficiency due to agents' misreport. We describe the system model and formulate the decomposable NUM Problems in Section \ref{NUMProm}.
We formally design the DeNUM Mechanism and the DeNUM Algorithm in Sections \ref{MechDeisn} and \ref{D-MEDEN}, respectively. In Section \ref{DyDeNUM}, we formally design the DyDeNUM Mechanism.
 In Section \ref{UPN}, we solve  a concrete example of user-provided network using the proposed DeNUM framework.
 Section \ref{Conclusion} concludes the paper.


\section{Literature review}\label{related}


\subsection{Mechanism Design for Network Function Virtualization}
	A group of literature related to our work is the mechanism design for network function virtualization (e.g. \cite{Virt1,Virt2,Virt3,Virt4,Virt5,NFV}), which is an important application of the network sharing paradigm.
	Specifically, in \cite{Virt1}, Fu and Kozat proposed to use  the Vickrey-Clarke-Groves (VCG) Mechanism \cite{Vic,Cla,Gro} to regulate the virtualized wireless resources. In \cite{Virt2}, Gu \textit{et al.} proposed an efficient auction for service chains in  the network function virtualization market. In \cite{Virt3}, Zhu and Hossain studied an interesting hierarchical auction for virtualization of 5G cellular networks. Du \textit{et. al.} in \cite{Virt4} proposed an auction traffic offloading based on software-defined network. Zhang \textit{et al.} in \cite{Virt5} considered a double auction for the virtual resource allocation of software-defined networks. Readers can refer to the survey \cite{NFV} for other related work.
	
	There are two main differences between our work and this group of literature. First,
	most works (e.g. \cite{Virt1,Virt2,Virt3,Virt4}) modeled the virtualized resources as \emph{indivisible goods} and used one-shot VCG-type mechanisms to achieve the network utility maximization. Reference \cite{Virt5} is an exception that considered the divisible virtualized resources but assumed that agents are \textit{price-takers} instead of strategic agents.
		In this paper, we consider the shared resources divisible, which can achieve more flexible  network sharing among agents. 
	Moreover, a one-shot VCG-type mechanism is not applicable here.
	 This is because (i) the one-shot VCG-type mechanism requires agents to report their entire utility functions, which incurs significant communication overheads due to the often high dimension information to fully describe the utility function, and (ii) it is impossible for a one-shot dominant-strategy allocation mechanism\footnote{In a dominant-strategy allocation mechanism, it is a dominant strategy for each agent to truthfully reveal her private information (independent of other agents' choices).} (such as a VCG-type mechanism) to achieve several properties including the network-utility maximization, budget balance, and individual rationality
	  at the same time \cite{publicgoods}. For instance, the VCG-type mechanism cannot achieve the budget balance; Ge and Berry  in \cite{Quant} proposed a dominant-strategy allocation mechanism by quantizing divisible goods but does not achieve the maximal network utility.
Finally,  the existing literature assumes that the constraint information is globally known.

\vspace{-0.2cm}
\subsection{Mechanism Design for the NUM Problems}

\subsubsection{Nash Mechanisms} 
Due to the above mentioned reasons, research efforts for divisible network resource allocation mainly prefer
 Nash mechanisms to the dominant-strategy allocation mechanisms (such as the aforementioned one-shot VCG mechanism).




There are many excellent works that proposed Nash mechanisms for specific allocations, such as 
the general flow control problems (e.g. \cite{VCGKelly,scalar,Jain2010,Multicast,Surrogate}), plug-in electric vehicles system (e.g. \cite{PEV}), power allocation and spectrum sharing problem (e.g. \cite{power}), networked public good \cite{sharma2012local}, networked private good \cite{learning}.
Sinha \textit{et al.} studied a relatively more general setting in \cite{General}, which is also a subclass of the problem that we study in this work. 
Only one work considered the private constraint information \cite{sharma2012local}, which focused on 
 the uncoupling local constraints for a specific setting instead of the more challenging coupling system constraints. 

\subsubsection{Dynamic Mechanisms} 

The aforementioned works considered one-shot mechanisms. References \cite{Dynamic1,Dynamic2} studied interesting dynamic mechanisms that dynamically implement Grove-like taxation \cite{Gro}, which motivate our DyDeNUM Mechansim. Different from \cite{Dynamic1,Dynamic2}, our DyDeNUM Mechanism is able to cope with the asymmetric constraint information and applicable to a more general class of the decomposable NUM Problems. 

We summarize the key features of the proposed DeNUM Mechanism  and the DyDeNUM Mechanism and the existing mechanisms for the NUM Problems in Table I.
\subsection{Distributed Algorithms for Nash Mechanisms}

Only a few studies focused on the distributed algorithms (dynamics) for the Nash mechanisms for network applications \cite{Surrogate,learning}. We cannot directly apply these algorithms in \cite{Surrogate,learning} in our context. This is because, for GNEs, the best response dynamics considered in \cite{learning} was proven to converge only in restrictive cases \cite{GNEP}, while  \cite{Surrogate} requires to solve a centralized optimization problem in each iteration, which is not available in the problems considered here.



\section{An Example of Inefficiency due to Misreports}\label{Illu}

To show that misreporting private constraint information can lead to efficiency loss, let us consider the following example.
\begin{example}
	Consider a network flow-control problem with one link provider and one end user (see \cite{Jain2010,General}). The link provider can allocate bandwidth $x_1$ to the user,  subject to a capacity constraint $c$. The user achieves a throughput $x_2$, which equals $Ax_1$ due to packet loss. We refer to $A\in(0,1]$ as the packet delivery ratio. 
	The link has a cost function of  $C(x_1)$ and the user has a utility function of  $U(x_2)$. The  corresponding NUM Problem is
	\vspace{-0.15cm}
		\begin{align}
		\max_{x_1,x_2}~U(x_2)-C(x_1)~~
		{\rm s.t.}\hspace{-0.2cm}\underbrace{x_2=Ax_1}_{\rm system~constraint}\hspace{-0.2cm},\overbrace{0\leq x_1\leq c}^{\rm local~constraint}. \label{IllNUM}
		\end{align}
\end{example}

	\begin{figure}[t]
		\centering
		\includegraphics[scale=0.35]{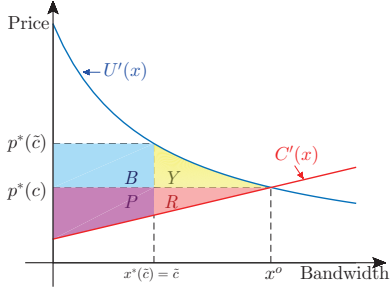}
		\vspace{-10pt}	
		\caption{The self-interested provider can under-report the capacity $c$ to improve his profit.  }
		\label{cdf}
\end{figure}

\begin{figure}[t]
		\centering
		\includegraphics[scale=0.35]{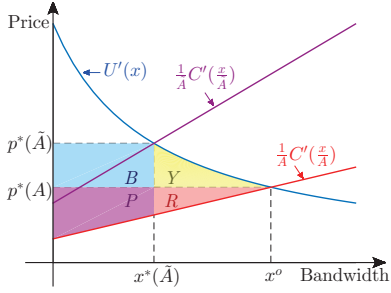}
		\vspace{-10pt}	
		\caption{The self-interested provider can under-report coefficient $A$ to improve his profit.  }
		\label{cdf2}
\end{figure}


Suppose that $A=1$ and is known by the system designer, while the parameter $c$ is the link provider's private information.
The link provider can misreport $c$, and it may be difficult for  the end user or the system designer  to verify.

 Let $c$ be sufficiently large. Consider the traditional mechanism mentioned (e.g. \cite{Jain2010,General}), which determines the throughput and a price per throughput $p$ based on the link provider's reported  value of  $\tilde{c}$. At a ``traditional" equilibrium, the price $p^*$ equals an optimal dual variable corresponding to the system constraint in \eqref{IllNUM} \cite{Jain2010,General}.
That is, the equilibrium $(x^*,p^*)$ satisfies
\vspace{-0.2cm}
	\begin{align}
	x^*(\tilde{c})&=\arg\max_{0\leq x\leq \tilde{c}}\left\{U(x)-C\left(x\right)\right\},\label{a11}\\
	p^*(\tilde{c})&=U'(x^*(\tilde{c})).\label{a22}
	\end{align}
Hence, the user's throughput is $x_2^\ast = x^*$. The link provider has allocation $x_1^\ast =  x^*$ and a profit of $x^* p^* - C (x^*)$. 
%

  As shown in Fig. \ref{cdf}, if the provider reports the true value of $c$, the mechanism's outcome $(x^*,p^*)$, as shown in \eqref{a11}-\eqref{a22}, is the coordinates of
	the intersection point of the two curves  $U'(x)$ and $C'(x)$, i.e., $(x^o, p^*(c))$. This leads to a profit of the link provider equal to the area of $R$ plus $P$, i.e., $x^*p^* - \int_0^{x^o} C'(x) dx$.   However, the provider can report a much smaller $\tilde{c}$, in which case $x^*=\tilde{c}$ according to \eqref{a11} and the price $p^*$ will increase. This
	results in a larger profit  of the link provider (which is equal to the area of $B$ plus $P$) than the one under truthful report. 	 On the other hand, it reduces the network utility  (which equals $U(x_2^\ast) - C(x_1^\ast)$) by the area of $Y$ plus $R$.
Therefore,  a strategic link provider will misreport $\tilde{c}$  to increase its profit, leading to the efficiency loss. 

We next show that misreporting $\tilde{A}$ can also lead to an efficiency loss.
At a ``traditional"  equilibrium, the price $p^*$ is equal to an optimal dual variable corresponding to the system constraint in \eqref{IllNUM} \cite{Jain2010,General}.
That is, there exists an equilibrium $(x^*,p^*)$ satisfying
\begin{align}
x^*(\tilde{A})&=\arg\max_{x\geq 0}\left\{U(x)-C\left(\frac{x}{\tilde{A}}\right)\right\},\label{a11111}\\
p^*(\tilde{A})&=U'(x^*(\tilde{A})).\label{a22222}
\end{align}
Hence, the user's throughput is $x_2^\ast = x^*$. The link provider has allocation $x_1^\ast =  x^*/A$ and a profit of $x^* p^* - C (x^*/A)$. 

As shown in Fig. \ref{cdf2}, 
if the provider reports the true value of $A$, the mechanism's outcome $(x^*,p^*)$, as shown in \eqref{a11111}-\eqref{a22222}, is the coordinates of
the intersection point of the two curves  $U'(x)$ and $C'(x/A)/A$, i.e., $(x^o, p^*(A))$. Similarly, this leads to a profit of the link provider equal to the area of $R$ plus $P$, i.e., $x^*p^* - 1/A\int_0^{x^o} C'(x/A) dx$.   However, the provider can report a smaller $\tilde{A}$, in which case $(x^*,p^*)$ becomes the coordinates of the intersection point of the two curves $U'(x)$ and $C'(x/\tilde{A})/\tilde{A}$. This
results in a higher price $p^*$ and thus a larger profit  of the link provider (which is equal to the area of $B$ plus $P$) than the one under truthful report. 	 On the other hand, it reduces the network utility  (which equals $U(x_2^\ast) - C(x_1^\ast)$) by the areas of $Y$ plus $R$, similar to the case in Fig. \ref{cdf}.	Misreporting both types of constraint information leading to an efficiency loss motivates this study.

\begin{table}[]
	\centering
\rev{	\caption{Notation}
	\vspace{-0.2cm}
	\label{notation}
	\begin{tabular}{|c|c|}
		\hline
		\hline
		Symbol              & Physical Meaning                                             \\ \hline
		$\mathcal{I}$         & Set of agents                              \\ \hline
		$\mathcal{N}_i$  & Set of all constraints that agent $i$'s action has influence on                            \\ \hline
		$\mathcal{I}_n$  & Set of all agents whose actions have influence on constraint $n$                           \\ \hline
		$\bs{x}_i$         & Action of agent $i$                                   \\ \hline
		$h_{i,n}(\cdot)$               &  Influence function of agent $i$ for system constraint $n$
		   \\ \hline
        $\trianglelefteq_n$               & Inequality sign $\leq$ or equals sign $=$ for system constraint $n$ \\ \hline
		$\mathcal{X}_i$               & Local constraint for agent $i$ \\ \hline
		$c_n$               & System constraint parameter for constraint $n$ \\ \hline
		$U_i(\cdot)$               & Utility for agent $i$ \\ \hline
	\end{tabular}}
	\vspace{-15pt}
\end{table}

\section{The Network Utility Maximization Problem} \label{NUMProm}
In this section, we introduce a network sharing framework of Network Utility Maximization (NUM) with decomposability structures. We first describe various  components of the model and then present the decomposable NUM (DeNUM) problem. 
%

\subsection{System Model} \label{Sysm}

A network-sharing NUM framework consists of agents, limited  resources characterized by several  constraints, and a global objective.

\subsubsection{Agents} We consider a networked system with a set $\mathcal{I}=\{1,...,I\}$ of agents. 
An agent can be either a service provider or a user, as we illustrated in Section \ref{Illu}. Each agent 
is rational and selfish, and hence aims to maximize her own benefit. 
  
\emph{Actions:} \rev{We use $\bs{x}_i=\{x_{i,l}, 1\leq l\leq L_i\}\in\mathbb{R}^{L_i}$ to denote agent $i$'s (allocative) \emph{action}, where $L_i$ is the dimension of agent $i$'s action. The value of $x_{i,l}$ captures consumption/sharing of one resource/service or a decision regarding one task.} Each agent $i$'s choice of $\bs{x}_i$ is subject to a \textit{local constraint} characterized by a feasible set $\mathcal{X}_i$, i.e., $\boldsymbol{x}_i \in \mathcal{X}_i$. Sharing/consuming no resource (or  making no action) is always a feasible choice, i.e.,  $\bs{0}\in\mathcal{X}_i$.

\emph{Utility Functions:} Each agent $i$ has a  \emph{utility function}  $U_i(\bs{x}_i)$, which denotes her  benefit (or the negative of her cost) as a function of her action $\bs{x}_i$.\footnote{The utility is allowed to be negative and decreasing in some dimensions.}

	\begin{remark}\label{R1}
	With a proper reformulation, our framework is applicable to the case where \textit{agent $i$'s utility $U_i(\cdot)$ is a function of other agents' actions}. Please refer to  Appendix \ref{DCU} for detailed explanations.
	\end{remark}

\subsubsection{System Constraints and Influences} Consider a set $\mathcal{N}=\{1,...,N\}$ of \textit{system constraints}. Each constraint $n$ couples a set $\mathcal{I}_n\subseteq\mathcal{I}$ of agents' actions. 
Let $h_{i,n}(\bs{x}_i)$ denote agent $i$'s \emph{influence} to the system constraint $n$. We consider the following additive form for
system constraint $n$:
\begin{align}
\sum_{i\in\mathcal{I}_n}h_{i,n}(\bs{x}_i)~\trianglelefteq_n~c_n,\label{Con}
\end{align}
\rev{where the symbol $\trianglelefteq_n$, associated with constraint $n$, represents either the equals sign $=$ or the inequality sign $\leq$};
$c_n$ denotes a system constraint parameter for constraint $n$. 
 Let $\mathcal{N}_i\triangleq\{n: i\in\mathcal{I}_n\}$ denote the set of constraints that  agent $i$'s action  has influence on.

	\begin{remark}
		 With a proper reformulation, our framework is also applicable to the case where \textit{the influence function couples all agents' actions}, as in Remark \ref{R1}. Please refer to  Appendix \ref{DCC} for detailed explanations.
	\end{remark}

An inequality constraint can capture, for example,  resource allocation budget constraints (such as capacity constraints). In this case, a positive (negative) $h_{i,n}(\bs{x}_i)$ indicates a certain amount of resource consumption (production).\footnote{It can also capture, for example, the interference to the networks, as we will discuss in Section \ref{UPN}.}
An equality constraint usually captures the \textit{balancing} constraints, 
such as a network flow balance constraint (e.g. \cite{comm,Open,Sensor})  and  a  market clearing constraint (e.g. \cite{smartgrid}). We assume $h_{i,n}(\bs{0})=0$, i.e., idleness leads to zero influence to the system.
Finally, the additive form in constraint \eqref{Con} is applicable in  a large range of networked applications (e.g. \cite{Sensor,comm,Open,smartgrid,Cloud}).

%

\subsubsection{Information Structure}We assume that $U_i(\cdot)$, $h_{i,n}(\cdot)$, and $\mathcal{X}_i$ are agent $i$'s \textit{private information} that may not be known by others. Though the structure of $h_{i,n}(\cdot)$ is private, we consider the following monitorability assumption:
\begin{assumption}[Monitorable Influence]\label{Assum1}
After agent $i$ performs her action $\bs{x}_i$, the network designer or some other agent in $\mathcal{I}_n$ can observe
 the output value of the function $h_{i,n}(\bs{x}_i)$.
\end{assumption}
For instance,
an agent or the network designer can observe the total amount of another agent's resource consumption/production (as illustrated by a concrete example in Appendix \ref{ExampleC}) or the interference generated by another agent (as illustrated in Section \ref{UPN}). Such an assumption is also motivated by the fact that the 5G network slice broker can obtain access to network monitoring measurements such as load and various key performance indicates \cite{NetSlice}.
We will further discuss how to eliminate the need of Assumption \ref{Assum1} in Section \ref{DyDeNUM}.

We assume that the system constraint parameter $c_n$ is globally known. However, by a proper reformulation (i.e., introducing auxiliary system and local constraints), our framework is also applicable to the case where some parameter $c_n$ is only known by some agent.  For detailed discussions, please refer to Appendix \ref{UC}.

\subsection{Network Utility Maximization Formulation:}
 The system designer is interested in solving the following NUM Problems with a decomposition structure defined as:
 \begin{definition}[DeNUM: Decomposable NUM]\label{D1}
 	A DeNUM Problem has the following structures:
 	\vspace{-0.1cm}
 \begin{subequations}\label{NUM}
 	\begin{align}
 	&\max_{\bs{x}}~\sum_{i\in\mathcal{I}}U_i(\bs{x}_i)\\
 	&~~{\rm s.t.}~\sum_{i\in\mathcal{I}_n}h_{i,n}(\bs{x}_i)~\trianglelefteq_n~c_n,~~\forall n\in\mathcal{N},\label{coupledconstraint}\\
 	&~~~~~~~\bs{x}_i\in\mathcal{X}_i,~~\forall i\in\mathcal{I}. \label{xi}
 	\end{align}
 \end{subequations}	
 \end{definition}


We adopt the following standard assumptions to ensure convexity, feasibility, and constraint regularity  of the problem:
\vspace{-0.2cm}
\begin{assumption}\label{Assum2} The DeNUM Problem satisfies: 
	\begin{enumerate}
		\item Each agent $i$'s utility function $U_i(\bs{x}_i)$ is continuous, strictly concave, and differentiable;\footnote{We do not assume monotonicity for the utility functions.}
		\item Each agent $i$'s influence functions $h_{i,n}(\bs{x}_i)$ are continuous and differentiable; 
$h_{i,n}(\bs{x}_i)$ is affine if $\trianglelefteq_n$ is $=$, and it
is convex if $\trianglelefteq_n$ is $\leq$;
	  \item The local constraint $\mathcal{X}_i$ is convex and compact;
	  \item The set  of all feasible actions $\tilde{\mathcal{X}}$ (where $\bs{x}\in\tilde{\mathcal{X}}$ iff all $\bs{x}_i$ satisfy \eqref{coupledconstraint}, \eqref{xi}) is non-empty.
%
	\end{enumerate}
\end{assumption}


Let ${\rm relint}(\mathcal{A})$ be the relative interior of the set $\mathcal{A}$ \cite[Ch. 2.1.3]{boyd2004convex}. We further adopt the following regularity  assumption.
\begin{assumption}[Slater's Condition]\label{Assum3}
	There exists a feasible solution $\tilde{\bs{x}}=\{\tilde{\bs{x}}_i\}_{i\in\mathcal{I}}$ such that $\tilde{\bs{x}}_i\in{\rm relint}(\mathcal{X}_i),~\forall i\in\mathcal{I}$ and
\begin{align}
\sum_{i\in\mathcal{I}_n}h_{i,n}(\tilde{\bs{x}}_i)\begin{cases}
< c_n,~{\rm for~every}\trianglelefteq_n {\rm representing \leq},\\
= c_n,~{\rm for~every}\trianglelefteq_n {\rm representing =}.
\end{cases}
\end{align}


\end{assumption}
 Assumptions \ref{Assum2} and \ref{Assum3} ensure the sufficiency and necessity of the Karush-Kuhn-Tucker (KKT) conditions to characterize the global optimal solution of the DeNUM Problem \cite{boyd2004convex}.

\subsection{Desirable Mechanism Properties}
It is well known that one can design a distributed algorithm to efficiently solve the DeNUM Problem provided agents are willing to follow the algorithm \cite{NUMTut}, as we will further show in Section \ref{ID}. However, such an approach is  not self-enforcing since strategic agents may  misreport information or tampering with the algorithms. Hence, we need to design economic mechanisms to induce network-utility maximizing equilibria, under which each agent will maximize her local payoff function that is determined by the mechanism. The economic mechanisms should satisfy the following three desirable economic properties and one technical property:

\begin{itemize}
	\item (E1) \textbf{ \textit{Efficiency}}: The mechanism induces an equilibrium that maximizes the network utility, i.e., achieves the optimal solution of the DeNUM Problem.
	\item (E2) \textit{\textbf{Individual Rationality}}: Every agent should not be worse off by participating in the mechanism.
	\item (E3) \textit{\textbf{Strong Budget Balance}}: The total payment from some agents equals the reimbursements to all remaining agents. That is, there is no need to inject or take money.
	\item (T1) \textit{\textbf{Dynamic Stability}}:
	 The mechanism admits a distributed iterative algorithm, along which the agents can achieve  the equilibrium.
\end{itemize}

 We will first design a Nash mechanism  to achieve the above properties (E1)-(E3)\footnote{As Section \ref{related} mentioned, we do not seek for another well-known property ``truthfulness'' (i.e., truthful report is a dominant strategy) since it is not achievable together with (E1)-(E3) and may incur significant overheads. 
 	 } as well as a corresponding distributed algorithm that achieves (T1). We then design a dynamic mechanism to achieve (E1), (E2), and (T1). It cannot achieve (E3) due to the induced VCG-type taxation.

\subsection{Conditions and Impossibility Results}\label{Condition}
In this subsection, we discuss the conditions where it is possible for a mechanism to achieve the properties (E1)-(E3). We then adopt the assumptions to rule out the impossible scenarios.

\subsubsection{Excludability}
We adopt the following assumption:
\begin{assumption}[Excludability]\label{Assum4}
The system designer can exclude each agent $i$ from the system, which is equivalent to 
the case where agent $i$ can only choose an action from the set:
\begin{align}
\mathcal{X}_i^{\rm Out}=\left\{\bs{x}_i: \bs{x}_i\in\mathcal{X}_i, ~ h_{i,n}(\bs{x}_i)~\trianglelefteq_n~0,~~\forall n\in\mathcal{N}\right\}.\label{Out}
\end{align}
\end{assumption}
To understand \eqref{Out}, recall that a positive $h_{i,n}(\bs{x}_i)$ can represent  consumption of a certain amount of resources. Intuitively, the excludability means that the system designer can prevent a non-paying agent from free-riding any network resource.

Fortunately, most resources (or services) in networked systems are excludable.\footnote{Specifically, bandwidth, cloud services, contents, and electricity are intrinsically excludable. Moreover, many seemingly non-excludable resources have been made excludable. For instance, licensed spectrum is excludable, since Federal Communications Commission (FCC) imposed exclusive rights for a licensed spectrum holder and provides legal protection against unauthorized usage. Exceptions are wireless power in wireless power transfer network \cite{WPT} and network security investments \cite{security}. } Moreover, almost all existing mechanisms implicitly adopted Assumption \ref{Assum4} (e.g. \cite{VCGKelly,scalar,Jain2010,Multicast,power,Surrogate,learning,PEV,sharma2012local,General}). The reason is that 
non-excludability is one of the greatest enemies preventing (E1)-(E3) from being possible (see \cite{saijo2010fundamental,security}).  Intuitively, if the agents can always access the resources, they may opt out of any mechanism to avoid possible payments.


\subsubsection{Impossibility Results}
We present conditions regarding parameters $\{c_n\}$ where no mechanism can achieve properties (E1)-(E3) for every DeNUM Problem.
\begin{proposition}\label{T1}
	 No mechanism that can achieve both (E2) and (E3) for all DeNUM Problems under one of the following conditions:
	 \begin{itemize}
	 	\item $c_n$ is negative for some $n$ such that $\trianglelefteq_n$ is $\leq$;
	 	\item $c_n$ is non-zero for some $n$ such that $\trianglelefteq_n$ is $=$.
	 \end{itemize}
\end{proposition}
\begin{IEEEproof}
Please see Appendix	\ref{ProofP1}.
\end{IEEEproof}

%
Intuitively, each agent $i$ can receive at least a utility of $U_i(\bs{0})$ after opting out of any mechanism.\footnote{This is because $h_{i,n}(\bs{0})=0,~\forall n\in\mathcal{N}_i$ and $\bs{0}\in\mathcal{X}_i$.} Under the conditions in Proposition \ref{T1}, 
 the achievable network utility may be so limited that someone must increase her payoff by opting out of any mechanism. 
 Therefore, we adopt the following assumption:
 \begin{assumption}[Feasibility of Null]\label{Assum5}
 	Agents' action profile $\bs{x}=\bs{0}$ is a feasible solution to the DeNUM Problems, i.e., $c_n\geq 0$ if $\trianglelefteq_n$ is $\leq$ and $c_n=0$ if $\trianglelefteq_n$ is $=$.
 \end{assumption}


%

\section{The DeNUM Mechanism}\label{MechDeisn}

In this section, we propose the DeNUM Mechanism. We first present the indirect decomposition method motivating the DeNUM Mechanism
and the key idea behind the DeNUM Mechanism. We then formally present the DeNUM Mechanism and show that it can achieve  (E1)-(E3).

\subsection{Indirect Problem Decomposition}\label{ID}

We first present the indirect (dual) decomposition \cite{NUMTut} that serves as a distributed (pure) optimization method for solving the DeNUM Problem when agents are obedient.
 We consider to relax the constraints in \eqref{Con} and then introduce auxiliary variables $\bs{y}_i=\{y_{i,n}\}$ for each agent $i$ and the corresponding auxiliary constraints. The DeNUM Problem is equivalent to the reformulated one
as shown in the following result: 
 \begin{lemma}[R-DeNUM: Reformulated Decomposable NUM]\label{L-R-NUM}
 	The DeNUM Problem defined in Definition \ref{D1} is equivalent to the following
R-DeNUM Problem:
\begin{subequations}\label{R-NUM}
\begin{align}
&\max_{\bs{x},\bs{y}}~\sum_{i\in\mathcal{I}}U_i(\bs{x}_i)\\
&~~{\rm s.t.}~~\sum_{i\in\mathcal{I}_n}y_{i,n}=c_n,&~~\forall n\in\mathcal{N},\label{t-constraint}\\
&~~~~~~~~h_{i,n}(\bs{x}_i)~\trianglelefteq_n~y_{i,n},&~~\forall i\in\mathcal{I}, n\in\mathcal{N}_i,\\
&~~~~~~~~\bs{x}_i\in\mathcal{X}_i,&~~\forall i\in\mathcal{I}.
\end{align}
\end{subequations}
 \end{lemma}
We can prove this lemma  by showing the equivalence of two problems' KKT conditions.
\rev{The key idea of this reformulation is to partition the system constraints in \eqref{Con} into several individual constraints and re-impose each of them to the corresponding agent. This constructs an \textit{indirect decomposition structure} \cite{NUMTut}. }

To see this, we relax the constraint in \eqref{t-constraint} and assign $\bs\lambda=\{\lambda_n\}_{n\in\mathcal{N}}$ to be the dual variables of it.  We can then formulate the corresponding Lagrangian, which can be further decomposed into $I$ locally solvable subproblems. 
That is, 
agent $i$'s local problem is: 
\vspace{-0.2cm}
\begin{subequations}\label{Indirect}
\begin{align}
g_i(\bs\lambda)\triangleq&\max_{\bs{x}_i\in\mathcal{X}_i,\bs{y}_i}~U_i(\bs{x}_i)-\sum_{n\in\mathcal{N}_i}\lambda_n \left(y_{i,n}-\frac{c_n}{|\mathcal{I}_n|}\right)\\
&~~{\rm s.t.}~h_{i,n}(\bs{x}_i)~\trianglelefteq_n~y_{i,n},~~\forall n\in\mathcal{N}_i,
\end{align}
\end{subequations}
where we define $g_i(\bs\lambda)$ as the \textit{local dual function}. 
At the higher layer, we obtain the optimal dual variable $\bs\lambda^o$ through solving a master (global) dual problem, given by
\begin{align}
\bs\lambda^o=\arg\min_{\bs\lambda}\sum_{i\in\mathcal{I}}g_i(\bs{\lambda}).\label{DP}
\end{align}
Substituting $\bs\lambda^o$ into \eqref{Indirect}, we will have the optimal primary variables $(\bs{x}_i^o,\bs{y}_i^o)$ for each agent $i$'s local problem. 

The above approach works only if agents are obedient. 
The agent rationality and selfishness motivate us to propose a mechanism to align
 strategic agents' interests to the above approach to solving the problem in \eqref{DP}.



\subsection{Key Ideas Behind  the DeNUM Mechanism}

Traditionally, a mechanism consists of a \textit{message space} and an \textit{outcome function} \cite{Implementation}, and each agent needs to  submit a message.
Such a mechanism is a tuple $\tilde{\Gamma}\triangleq(\mathcal{M},O)$, where the set $\mathcal{M}$ is the space from which  agents choose the messages $\bs{m}$; the outcome maps their message to the action agents should take $\hat{\bs{x}}(\bs{m})\triangleq\{\hat{\bs{x}}_i(\bs{m})\}_{i\in\mathcal{I}}$ and agents' payments 
 $\bs{\Pi}(\bs{m})\triangleq\{\Pi_i(\bs{m})\}_{i\in\mathcal{I}}$, i.e., $O(\bs{m})\triangleq(\hat{\bs{x}}(\bs{m}),\bs{\Pi}(\bs{m}))$. 
However, to design a mechanism $\tilde\Gamma$ with constraint information asymmetries, we need to find a mapping $O(\bs{m})$ that not only solves the DeNUM Problem but also incentivizes agents to reveal their private information, which is challenging.

In this paper, we propose a new mechanism framework, where a mechanism does not directly determine the allocation outcome.
Instead, each agent \textit{simultaneously submits a message and selects her allocative action $\bs{x}_i$ from an action set determined by the mechanisms}.\footnote{Our proposed mechanism framework generalizes $\tilde{\Gamma}$,
 which corresponds to the special case of our proposed framework where each agent's action space only contains one element (i.e., $\mathcal{T(\bs{m})}=\{\hat{\bs{x}}(\bs{m})\}$).  }
Specifically, the considered mechanism is a tuple $\Gamma\triangleq(\mathcal{M},\mathcal{T}(\boldsymbol{m}),\bs{\Pi}(\bs{m}))$:
the set $\mathcal{M}$ is the message space. The set $\mathcal{T(\bs{m})}\triangleq\{\mathcal{T}_i(\bs{m})\subseteq \mathcal{X}_i\}_{i\in\mathcal{I}}$ characterizes each agent's action space $\bs{x}_i$. A key feature (and challenge of the analysis later on) is that $\mathcal{T}(\boldsymbol{m})$  depends not only on $\mathcal{X}_i$ but also on some (unspecified) constraint determined by 
messages $\boldsymbol{m}$ announced by agents (which results in coupling among agents). Function $\bs{\Pi}(\bs{m})$ describes agents' payments  (also called taxes in the proposed framework).



The advantages of such a mechanism framework are two-fold. First, the computation of the allocation outcome is distributed and performed locally by agents. Second, by carefully designing a mechanism, only the agents need to utilize the private (utility and constraint) information for solving their own local problems. This eliminates the necessity for revealing agents' constraint information through a mechanism.

\subsection{DeNUM Mechanism and its Induced Game}

\subsubsection{Formal Mechanism Design} We introduce the DeNUM Mechanism which describes the message space $\mathcal{M}$, 
budgets constraining agents' actions $\mathcal{T(\bs{m})}$,
 and their taxes $\bs\Pi(\bs{m})$.

\hspace{-0.5cm}	\qed
	\vspace{-0.3cm}
\begin{mechanism}[DeNUM] The DeNUM Mechanism consists of the following components:
	\vspace{-0.35cm}
	
\hspace{-0.5cm}	\qed
	\begin{itemize}
		\vspace{-0.15cm}
		\item \textbf{The message space $\mathcal{M}=\times_{i\in\mathcal{I}}\mathcal{M}_i$}: Each agent $i\in\mathcal{I}$ submits a message $\bs{m}_{i}=\{m_{i,n}\}_{n\in\mathcal{N}_i}\in\mathcal{M}_i\triangleq\mathbb{R}^{2\times|\mathcal{N}_i|}$ to the system designer: \footnote{Note that we allow the price proposal $p_{i,n}$ to be negative, in which case it represents a proposed reimbursement per unit of allocated budget. }
		\vspace{-0.2cm}
		\begin{align}
		m_{i,n}=(p_{i,n},\tau_{i,n}),
		\end{align}	
		where $p_{i,n}$ and $\tau_{i,n}$ denote agent $i$'s \textbf{price proposal} and \textbf{budget proposal}, respectively. We denote all agents' message profile as $\bs{m}=\{\bs{m}_i\}_{i\in\mathcal{I}}$.
		\item \textbf{Imposed Constraints}:
		For the action $\bs{x}_i$ for agent $i\in\mathcal{I}$,  
		  the system designer imposes an additional budget constraint on the agent's  influence $h_{i,n}(\bs{x}_i)$, denoted  by
		\begin{align}
		h_{i,n}(\bs{x}_i)\trianglelefteq_n  t_{i,n}(\bs{\tau}_n),~ n\in\mathcal{N}_i,\label{constraint1}
		\end{align}
		where $\bs{\tau}_n=\{\tau_{i,n}\}_{i\in\mathcal{I}_n}$ and $t_{i,n}$ is agent $i$'s \textbf{budget} associated with system constraint $n$, denoted by
		\vspace{-0.1cm}
		\begin{align}
		t_{i,n}(\bs{\tau}_n)&=\tau_{i,n}-\frac{\sum_{j\in\mathcal{I}_n} \tau_{j,n}-c_n}{|\mathcal{I}_n|}. \label{constraint2}
		\end{align}
		\item \textbf{Taxation $\bs\Pi$}: For each system constraint $n\in\mathcal{N}_i$, each agent $i$ pays a tax of \footnote{A negative tax corresponds to  a reimbursement from the system designer.}
		\vspace{-0.1cm}
		\begin{align}
		&\pi_{i,n}(\bs{m}_n)\label{payment}\\
		=&\underbrace{p_{\omega(n,i+1),n}\left(t_{i,n}(\bs{\tau}_n)-\frac{c_n}{|\mathcal{I}_n|}\right)}_{\rm payment}
		+\underbrace{(p_{i,n}-p_{\omega(n,i+1),n})^2}_{\rm penalty},\nonumber
		\end{align}
			where $\bs{m}_n=\{m_{i,n}\}_{i\in\mathcal{I}_n}$. Here $\omega(n,i+1)$ denotes the circular neighbor of agent $i$ on constraint $n$. More specifically,  suppose $i$ is the $\upsilon$-th smallest index in $\mathcal{I}_n$, then\footnote{For example, $\omega(n,3+1)=5$ when $\mathcal{I}_n=\{2,3,5\}$.}
			\begin{align}
			\hspace{-0.6cm}\omega(n,i+1)\triangleq\begin{cases}
			\text{the $(\upsilon+1)$-th smallest index in $\mathcal{I}_n$, if $\upsilon\neq|\mathcal{I}_n|$}, \nonumber\\
			\text{the smallest index in $\mathcal{I}_n$,  otherwise}.
			\end{cases}
			\end{align}
			Agent $i$'s total tax is
			\vspace{-0.1cm}
				\begin{align}
				\Pi_{i}(\bs{m})=& \sum_{n\in\mathcal{N}_i}\pi_{i,n}(\bs{m}_n).\label{Payment}
				\end{align}
	
		\end{itemize}			
			\vspace{-0.3cm}
	\end{mechanism}		
\hspace{-0.5cm}	\qed

In our DeNUM Mechanism, each agent should simultaneously submit two types of messages  (price and budget) and decide her action $\bs{x}_i$. For each system constraint $n$, proposal $\tau_{i,n}$ denotes the budget $t_{i,n}$ that agent $i$  demands; $p_{i,n}$ denotes the price that agent $i$ is willing to pay. Both $\mathcal{X}_i$ and the constraints specified by \eqref{constraint1}-\eqref{constraint2} constrain agent $i$'s possible  strategy. Finally, each agent pays a tax \eqref{payment}-\eqref{Payment} associated with other agents' price proposals and her own budgets.



Note that constraints in \eqref{constraint1}-\eqref{constraint2} can be either ``hard'' physical constraints or ``soft''  contractual constraints. In the latter case, each agent is  still able to violate the constraints, but such violation is detectable by
 comparing the output of the function $h_{i,n}(\bs{x}_i)$ (by Assumption 1) and $t_{i,n}(\bs\tau_n)$, without requiring the knowledge of the exact forms of $h_{i,n}(\bs{x}_i)$.
Note that agents are willing to monitor each other on behalf of the system designer and report any violator. This is because
 one agent's violating action will harm the benefit of another, since the latter may not access the whole budget as promised in \eqref{constraint1}.
%
 Therefore, as far as the mechanism is concerned, we assume that the constraints in \eqref{constraint1}-\eqref{constraint2} are ``hard'' and inviolable.




\rev{The taxation for each agent $i$ in \eqref{payment} consists of a payment term for her budget and a penalty term. 
The payment term regulates agents' demands of the budget in such a way that each agent's payoff has a similar structure to the objective in \eqref{Indirect}.
The penalty term is motivated by \cite{hurwicz1979outcome}, which penalizes price proposal deviations to incentivize similar price proposals and is designed to become zero at the induced equilibrium.}


\subsubsection{DeNUM Game}
The above DeNUM Mechanism induces a DeNUM Game where each agent simultaneously decides  $\bs{m}_i=\{m_{i,n}\}_{n\in\mathcal{N}}$ and $\bs{x}_i$, aiming to maximize her utility minus her tax in \eqref{Payment} and considering other agents' decisions:
\begin{game} (Induced by the DeNUM Mechanism)
	\begin{itemize}
		\item Players: all agents in $\mathcal{I}$;
		\item Strategy Space: for agent $i\in{\mathcal{I}}$, her strategy space is $(\bs{x}_i,\bs{m}_i)\in \mathcal{S}_i(\bs{\tau}_{-i})$,
where\footnote{The strategy space in \eqref{SS} for each agent is always non-empty. This is because $h_{i,n}(\bs{0})=0$ and each agent $i$ can always submit an appropriate $\tau_{i,n}$ to ensure $t_{i,n}(\bs{\tau}_i; \bs{\tau}_{-i})=0$. Therefore, agents can always ensure the  feasibility of \eqref{SS}, regardless of other agents' message $\bs{m}_{-i}$.}
\begin{align}
\mathcal{S}_i(\bs{\tau}_{-i})\triangleq\left\{(\bs{x}_{i},\bs{m}_i): \bs{x}_i\in\mathcal{X}_i, \bs{m}_i\in\mathbb{R}^{2\times |\mathcal{N}_i|}, \right.\nonumber\\
\left.  h_{i,n}(\bs{x}_i)\trianglelefteq_n t_{i,n}(\bs{\tau}_i; \bs{\tau}_{-i})~,\forall n\in\mathcal{N}_i \right\};\label{SS}
\end{align}
		\item (Quasi-linear) payoff function $J_i(\cdot)$: each agent $i$ has a payoff function
	\begin{align}
	J_i(\bs{x}_i,\bs{m})\triangleq
	U_i(\bs{x}_i)-\Pi_i(\bs{m}).
	\end{align}
	\end{itemize}
\end{game}

Different from the traditional mechanisms \cite{Implementation}, the DeNUM Mechanism induces a game where each agent's strategy includes $\bs{m}_i$ and
$\bs{x}_i$ chosen from coupled strategy spaces.

\subsubsection{Generalized Nash Equilibrium}
The game-theoretic solution concept for the DeNUM Game  is the generalized Nash equilibrium (GNE) \cite{GNEP}.\footnote{\rev{The standard GNE (or an NE) usually stands for a solution concept for a  game with complete information, which is not the case here. 
	Instead, we adopt the common interpretation in the literature of Nash mechanisms (see \cite{Multicast,Surrogate,General,learning}). That is, a GNE
	is a  ``stationary'' point of some strategy updating processes (to be described in Section \ref{SV}) that possesses the equilibrium property in \eqref{NE}.}} This concept generalizes the traditional NE since agent strategies impact
not only other agents' payoffs but also other agents' strategy space.  
\begin{definition}[Generalized Nash Equilibrium (GNE)]
	A GNE of the DeNUM Game is a strategy profile $(\bs{x}^*,\bs{m}^*)$ such that for every agent $i\in\mathcal{I}$ and every strategy $(\bs{x}_{i},\bs{m}_i)\in\mathcal{S}_i(\bs{\tau}_{-i}^*)$,
	\begin{align}
	&J_i(\bs{x}_{i}^*,\bs{m}_i^*;\bs{m}_{-i}^*)\geq J_i(\bs{x}_{i},\bs{m}_i;\bs{m}_{-i}^*), \label{NE}
	\end{align}
	where $\bs{m}_{-i}^*\triangleq \{\bs{m}_{j}^*\}_{j\neq i}$ is the GNE strategy profile of all
	other agents except agent $i$.
\end{definition}

\subsection{GNE Analysis}

\subsubsection{GNE Price Proposals}

For each agent $i$, her price proposal only affects the penalty term in \eqref{payment}. We can verify  that each agent $i$ will always choose  $p_{i,n}=p_{\omega(n,i+1),n}$ for every system constraint $n\in\mathcal{N}_i$ to minimize the penalty.
This leads to the following result. 
\begin{lemma}[Common Price Proposals]\label{L22}
The GNE price proposals satisfy that, for each system constraint $n$, 
\begin{align}
p_n^*\triangleq p_{i,n}^*=p_{j,n}^*,~\forall i,j\in\mathcal{I}_n. \label{NEprice}
\end{align}
\end{lemma}

By Lemma \ref{L22}, since every agent submits her price proposals according to \eqref{NEprice}, every penalty term in \eqref{payment} is zero. In addition, the  budgets determined in \eqref{constraint2} ensure that $\sum_{i\in\mathcal{I}_n}t_{i,n}(\bs{\tau}_n)=c_n$ for every $\bs{\tau}_n$. It follows that
\begin{proposition}[Budget Balance]\label{P3}
The DeNUM Mechanism satisfies the budget balance (E3), i.e., $\sum_{i\in\mathcal{I}}\Pi_{i}(\bs{m}^*)=0.$
\end{proposition}

\subsubsection{Agent Payoff Maximization}	


By Lemma \ref{L22} and \eqref{NE},  agents achieve a GNE if, under the properly selected common price proposals $\{p_n^*\}_{n\in\mathcal{N}}$, each agent $i$ solves 
the following convex Agent Payoff Maximization (APM) Problem: 
	\begin{align}\label{APM}
	\max_{\bs{x}_i\in\mathcal{X}_i,\bs\tau_i}~U_i(\bs{x}_i)-\sum_{n\in\mathcal{N}_i}p_n^{*}t_{i,n}(\bs\tau)~~~{\rm s.t.}~\eqref{constraint1}.
	\end{align}
  In other words, the KKT conditions of the APM Problem determine both $\{p_n^*\}_{n\in\mathcal{N}}$ and $\{(\bs{x}_i^*,\bs{\tau}_i^*)\}_{i\in\mathcal{I}}$.
  
	The APM Problem has a similar structure to the local problem in \eqref{Indirect}.
	However, different from \eqref{Indirect}, each agent self-enforcingly solves the APM Problem because it leads to her maximal payoff at a GNE.
	 In other words, the mechanism aligns each agent's interest with the decomposed optimization problem in \eqref{Indirect}. 
	 Moreover, only agent $i$ solving her APM Problem requires the knowledge of $U_i(\cdot)$, $h_{i,n}(\cdot)$, and $\mathcal{X}_i$. This resolves our main issue of information asymmetries and leads to the following results.
\begin{theorem}[Existence, Efficiency, and Full Implementation]\label{T3} 
 There exists at least one GNE in the DeNUM Game. When Assumptions \ref{Assum1}-\ref{Assum3} hold,
every GNE leads to the optimal solution to the DeNUM Problem (E1).
\end{theorem}
    \rev{\noindent \textit{Proof Sketch}: For any optimal solution $(\bs{x}^o,\bs\lambda^o)$ of the R-DeNUM Problem, a strategy profile $(\bs{x}^*, \bs{m}^*)$ that satisfies the following property is always a GNE: $\forall (i,n)\in\{(i,n): i\in\mathcal{I}_n\}$,
    	$\bs{x}_i^*=\bs{x}_i^o,~\tau_{i,n}^*=h_{i,n}(\bs{x}_i^o),$ and $p_{i,n}^*=\lambda_n^o$. This proves the existence of the GNE.
    	Due to the similarity of the structure between the APM Problem and the problem in \eqref{Indirect}, we can show the equivalence between the KKT conditions of the DeNUM Problem and those of all agents' APM Problems combined. Every GNE is thus a network-utility maximum. 
    	
    	Please refer to Appendix	\ref{ProofT1} for the complete proof.
    	\hfill\qedsymbol}


	
%
\begin{theorem}[Individual Rationality]\label{T4}
When Assumptions \ref{Assum1}-\ref{Assum5} hold, the DeNUM Mechanism is individually rational (E2).
\end{theorem}
    
\rev{\noindent \textit{Proof Sketch}: By Assumption \ref{Assum3}, if agent $i$ chooses not to participate in the mechanism, her maximal payoff is $\max_{\bs{x}_i\in\mathcal{X}_i^{\rm Out}}U_i(\bs{x}_i)$. If agent $i$ chooses to participate, she can always submit a message $\hat{\bs{m}}_i=(\hat{\bs\tau}_i,{\bs p}_i^*)$ where $\hat{\tau}_{i,n}=\frac{\sum_{i\neq j} \tau_{j,n}^*-c_n}{I-1}$, which leads to $t_{i,n}=0$. Therefore, her maximal payoff at a GNE is at least $\max_{\bs{x}_i\in\mathcal{X}_i^{\rm Out}}U_i(\bs{x}_i)+\sum_{n\in\mathcal{N}_i}p_{n}^*c_n/|\mathcal{I}_n|$. We then show that the term $\sum_{n\in\mathcal{N}_i}p_{n}^*c_n/|\mathcal{I}_n|$ is always non-negative at a GNE.
	
	Please refer to Appendix	\ref{ProofT2} for the complete proof.
\hfill\qedsymbol}
    

In a nutshell, the DeNUM Mechanism achieves (E1)-(E3) when  Assumptions \ref{Assum1}-\ref{Assum5} hold.

\section{Distributed Algorithm to Achieve the GNE}\label{D-MEDEN}

In the DeNUM Game, each agent does not directly know her GNE strategy satisfying \eqref{NE} due to private information regarding utilities and constraints.
 Hence, we propose the DeNUM algorithm for agents to distributively update their strategy and attain a GNE.\footnote{\rev{Due to the possibility of multiple primal solutions to the DeNUM Problem and multiple dual solutions to the dual problem in (10), different GNEs may lead to different individual payoffs
 		and hence
 		different agents might have different preferences in terms of different equilibria. 	
 		Note that if different agents choose to play their corresponding equilibrium strategies corresponding to different equilibria, then the overall strategy profile of all players may not be an equilibrium.  
 		Therefore, reaching one GNE relies on agents' consensus by  following the DeNUM Algorithm.}} We then prove its convergence.


\subsection{The Iterative DeNUM Algorithm}\label{SV}


	
    	Algorithm \ref{algo1} shows the proposed iterative DeNUM Algorithm for agents to distributively compute their GNE, with the key steps explained in the following.
	Each agent $i\in\mathcal{I}$ first  initializes her message $\bs{m}_i[0]\in\mathbb{R}^{2\times |\mathcal{N}_i|}$ (line \ref{l2}). The algorithm iteratively computes each agent's message and action until convergence (lines \ref{line4}-\ref{line188}). For each iteration, agents update their messages and actions in a Gauss-Seidel fashion (lines \ref{line6}-\ref{line14}). That is, we divide one iteration into $\mathcal{I}$ sub-iterations (line \ref{line6}). In each sub-iteration $i$, only agent $i$ updates her messages and actions,  whereas the other agents keep theirs fixed.

		\begin{algorithm}[tb]
			\SetAlgoLined
				\caption{DeNUM Algorithm: Distributive Computation of the GNE}\label{algo1}
				Initialize the iteration index  $k\leftarrow 0$\;
				Each agent $i \in\mathcal{I}$ randomly initializes $\{p_{\omega(i-1,n),n}[0]\}_{n\in\mathcal{N}_i}$ and the system designer chooses the stopping criterion $\epsilon$\label{l2}\;
				${\rm conv\_flag}\leftarrow 0$ $\#$ \textit{initialize the convergence flag}\; 
				\While{${\rm conv\_flag}= 0$ \label{line4}}{
					Set $k\leftarrow k+1$\; 
					\For{Each sub-iteration $i\in\mathcal{I}$}{\label{line6}
						Each agent $i$ updates $(\bs{x}_i[k], \bs{\tau}_i[k], \bs{p}_i[k])$ by \eqref{update}\; \label{line7}   
						\eIf{$||\bs{p}_i[k]-\bs{p}_i[k-1]||_2<\epsilon||\bs{p}_{i}[k-1]||_2$\label{lineif}}{
							${\rm conv\_count}\leftarrow {\rm conv\_count}+1$\;
						}{ ${\rm conv\_count}\leftarrow 0$\label{l3}\;}
						Each agent $i$ sends ${p}_{i,n}[k]$ to agent $\omega(i+1,n)$ for every $n\in\mathcal{N}_i$ and ${\rm conv\_count}$ to agent $i+1$\; \label{line8}
					}\label{line14}
					\If{${\rm conv\_count}=I$ \label{line15}}{
						Set ${\rm conv\_flag}\leftarrow 1$ and broadcasts it\label{l16}\;
					}		  	
				}\label{line188}
				Each agent $i$ submits message $\bs{m}_{i}$ to the system designer\; \label{line18}
				The system designer computes $(\bs{t}_i(\bs{m}), {\Pi}_i(\bs{m}))$ according to \eqref{constraint2}-\eqref{Payment} and sends them to every agent $i$\; \label{line19}
			%
		\end{algorithm}

	In particular, in each sub-iteration $i$, each agent $i$ updates $(\bs{x}_i[k], \bs{\tau}_i[k], \bs{p}_i[k])$ (in line \ref{line7}) according to
	\vspace{-0.15cm}
	         \begin{subequations}\label{update}
			 \begin{align}
			 &(\bs{x}_i[k], \bs{\tau}_i[k])\in\arg\max_{\substack{\bs{x}_i\in\mathcal{X}_i\\ \bs{t}_i}}~\left\{U_i\left(\bs{x}_i\right)-\sum_{n\in\mathcal{N}_i}p_{\omega(i-1,n),n}[k]t_{i,n}\right\}\nonumber\nonumber\\
			 &{\rm s.t.}~~~~~~~~~h_{i,n}(\bs{x}_i)\trianglelefteq_n t_{i,n},~t_{i,n}\leq t_{i,n}^{\rm up},~\forall n\in\mathcal{N}_i,\label{update1}
			 \end{align}
			 \begin{align}
			 &p_{i,n}[k]=\bar{p}_{\omega(i-1,n),n}[k]+\alpha[k]\left(\tau_{i,n}[k]-\frac{c_n}{|\mathcal{I}_n|}\right), ~\forall n\in\mathcal{N}_i, \label{update2}
			 \end{align}
			 \end{subequations}
where $\bar{p}_{\omega(i-1,n),n}[k]$ satisfies
	\begin{align}\label{mostrecent}
	&\bar{p}_{\omega(i-1,n),n}[k]\\
	\triangleq& \begin{cases}
	p_{\omega(i-1,n),n}[k-1],&~{\rm if~}i~{\rm is~the~smallest~index~in}~\mathcal{I}_n,\\
	p_{\omega(i-1,n),n}[k],&~{\rm otherwise},
	\end{cases}\nonumber
	\end{align} 			 
and $\alpha[k]$ is a diminishing step size, given by $\alpha[k]=(1+\beta)/(k+\beta)$ for some non-negative constant $\beta$. The explanation is as follows. First, each agent $i$ maximizes  her payoff function in \eqref{update1}, expecting that there is no penalty term in \eqref{payment} and her budgets equals to her budget proposals (i.e., $\bs{t}_i=\bs{\tau}_i$). The algorithm will satisfy  these two expectations when it converges as we will show. Second, each agent $i$ sets the additional upper bound $t_{i,n}^{\rm up}$ in \eqref{update1} by
			 \begin{align}
			 t_{i,n}^{\rm up}=\max_{\bs{x}_i\in\mathcal{X}_i}h_{i,n}(\bs{x}_i),~~~\forall (i,n)\in\{(i,n):i\in\mathcal{I}_n\},
			 \end{align} 
			 which ensures that the submitted $\{\bs{\tau}_i[k]\}$ are bounded. Such an upper bound always exists due to the compactness of the set $\mathcal{X}_i$.
Third, in \eqref{update2}, each agent updates  her price proposals to resemble others' most recent updated price proposals to reduce the penalty in \eqref{payment}.
			 

%

    Finally, each agent checks
    the termination criterion (line 
    \ref{l16}). The termination happens
    if
    the relative changes of all agents' price proposals in any continuous $I$ sub-iterations are small  enough. When the algorithm converges, agents submit their messages to the system designer to compute their imposed constraints and taxes (lines \ref{line18}-\ref{line19}). 
    
    Regarding  the complexity  of the algorithm, the system designer has very few computation and communication overheads  ($\mathcal{O}(IN)$ in line \ref{line19}). Each agent also encounters a small communication overhead  ($\mathcal{O}(N)$ per iteration in line \ref{line8}).

\subsection{Convergence Analysis}

To prove the convergence  of Algorithm \ref{algo1}, we first establish the connection between a GNE and the  optimal primal-dual solution of the R-DeNUM Problem in the following:
\begin{theorem}[Equivalence]\label{D2}
 When  Assumptions \ref{Assum1}-\ref{Assum3} hold, for a given optimal primal-dual solution $(\bs{x}^o,\bs{y}^o,\bs{\lambda}^o)$ to the R-DeNUM Problem in \eqref{R-NUM}, the following strategy profile $(\bs{x}^*,\bs{\tau}^*,\bs{p}^*)$ is a GNE: for all $(i,n)\in\{(i,n):i\in\mathcal{I}_n\}$,
	\begin{align}
	x^*_{i,n}=x^o_{i,n},~\tau_{i,n}^*=y_{i,n}^o,~p_{i,n}^*=p_n^*=\lambda_n^o, \label{critical}
	\end{align}
where $p_n^*$ is the common price proposal in Lemma \ref{L22}.
\end{theorem}
The proof of Theorem \ref{D2} involves exploiting its KKT conditions of the R-DeNUM Problem in \eqref{R-NUM} and those of the APM Problems in \eqref{APM}.

The significance of Theorem \ref{D2} is two-fold. First, it provides a new interpretation of the messages of the DeNUM Mechanism. Specifically, the budget proposals $\bs{\tau}_i$ play a role of the auxiliary variables $\bs{y}_{i}$ while each comment price proposal $p_n^*$ plays a role of a dual variable $\lambda_n$. Second, Theorem \ref{D2} suggests that any distributed algorithm updating $(\bs{x},\bs{m})$ to a primal-dual solution also converges to a GNE of the DeNUM Game.\footnote{Therefore, in addition to Algorithm \ref{algo1}, we can also adopt other algorithms (e.g. those from
\cite{Nedic2009} and \cite{Lobel2011}). Different from  Algorithm \ref{algo1}, they operate in the Jacobi (concurrent) and asynchronous fashion.} We prove the convergence next:
\begin{proposition}\label{T5}
When Assumptions \ref{Assum1}-\ref{Assum3} hold, the DeNUM Algorithm converges to a GNE.
\end{proposition}
\begin{IEEEproof}
	Please refer to Appendix \ref{ProofP3}.
\end{IEEEproof}
The proof of the proposition involves showing that the DeNUM Algorithm \ref{algo1} is an incremental subgradient based algorithm \cite{Incremental} which updates a dual variable incrementally in \eqref{update2}. Such an algorithm is convergent when $\alpha[k]$ is a diminishing step size and $\bs\tau_i[k]$ is bounded \cite{Incremental}.

\section{The Dynamic DeNUM Mechanism}\label{DyDeNUM}

The success of the DeNUM Mechanism relies on Assumption \ref{Assum1}, i.e., the output values of each agent's influence functions $h_{i,n}(\bs{x}_i)$ are monitorable. In this section, we propose the DyDeNUM Mechanism, a dynamic mechanism that can achieve (E1)-(E2) and (T1) even if the influence functions are not monitorable  (hence without Assumption 1). The tradeoff is that the DyDeNUM Mechanism is not guaranteed to satisfy  the budget balance (E3).

\subsection{The DyDeNUM Mechanism}

We  formally introduce the DyDeNUM Mechanism in Mechanism \ref{M2}. Different from the DeNUM Mechanism, the DyDeNUM Mechanism is executed in a dynamic fashion with the key steps introduced in the following.

\hspace{-0.6cm}	\qed
\vspace{-0.3cm}

\begin{mechanism}\label{M2} Dynamic DeNUM Mechanism (DyDeNUM)
	
	\vspace{-0.28cm}
	\hspace{-0.6cm}	\qed
	
	\vspace{-0.18cm}
	\begin{itemize}	
		\item \textbf{Initialization}: The system designer initializes taxes for agents $\{\Pi_i[0]\}_{i\in\mathcal{I}}$. Each agent initializes her price proposals $\bs{p}_{i}[0]=\{p_{i,n}[0]=C\}_{n\in\mathcal{N}_i}$ for some common constant $C>0$.
		\item For each iteration $k\in\mathbb{N}$,
		 \begin{itemize}
		 	\item For each sub-iteration $i\in\mathcal{I}$, each agent $i$ updates her message $\bs{m}_i[k]$:
			\begin{itemize}
				\item \textbf{Demand update}: 
				Agent $i\in\mathcal{I}$ reports her desired \textbf{demand} $\bs{d}_i[k]\in\mathbb{R}^{L_i}$ to the system designer.
				\item  \textbf{Price proposal update}: Agent $i$ updates her price proposal $p_{i,n}[k]$ and report it to her neighbor $\omega(i+1,n)$, for each resource $n\in\mathcal{N}_i$.
			
				\item \textbf{Marginal utility report:} Agent $i$ sends a reported marginal utility ${\nabla u_i}[k]\in\mathbb{R}^{L_i}$
			to the system designer.
					\end{itemize}
				\item \textbf{Taxation} $\bs{\Pi}[k]$: The system designer updates the taxation for each agent $i$, given by
				\begin{align}
				\hspace{-1cm}\Pi_{i}[k]=\Pi_{i}[k-1]+\sum_{j\in\mathcal{I}\slash\{i\}}\nabla u_j[k]^T(\bs{d}_{j}[k-1]-\bs{d}_{j}[k]).\label{Asu}
				\end{align}
			\end{itemize}

	\end{itemize}
\end{mechanism}
\vspace{-0.15cm}  	\qed

Each iteration of the DyDeNUM Mechanism consists of $I$ sub-iterations (similar to the DeNUM Algorithm). In each sub-iteration $i$, each agent $i$ should sequentially (in a Gauss-Seidel manner) submit three types of messages including price proposals, demand, and marginal utility. At the end of each iteration, the system designer updates each agent's tax in \eqref{Asu} based on other agents' reported price proposals and demands.


Given the tax in \eqref{Asu} and all other agents' messages $\bs{m}_{-i}$, each agent $i$ aims at maximizing her long-term average payoff:
\begin{subequations}
\begin{align}
&\max_{\{\bs{m}_i[k]\}_{k\in\mathbb{N}}}~\lim_{K\rightarrow \infty}\frac{1}{K}\left[\sum_{k=1}^KU_i(\bs{d}_i[k])-\Pi_i[k]\right]\\
&~~~~{\rm s.t.}~~~~~~~\bs{d}_i[k]\in\mathcal{X}_i,~\forall k\in\mathbb{N}.
\end{align}
\end{subequations}


We will show that each agent is interested in updating and reporting the message $\{\bs{m}_i[k]\}_{k\in\mathbb{N}}$ in the following way: for each iteration $k\in\mathbb{N}$,
\begin{align}
&\bs{d}_{i}[k]=\arg\max_{\bs{x}_i\in\mathcal{X}_i}\left[U_{i}(\bs{x}_i)-\sum_{n\in\mathcal{N}_i}\bar{p}_{\omega(i-1,n),n}[k] h_{i,n}(\bs{x}_i)\right],\label{Update1}~\\
&{p}_{i,n}[k]
=\label{Update2}\\
&\begin{cases}\bar{p}_{\omega(i-1,n),n}[k]+\alpha[k]\left( h_{i,n}(\bs{d}_i[k])-\frac{c_n}{|\mathcal{I}_n|}\right),~~~~{\rm if}~\trianglelefteq_n {\rm is~}=,\\\left[\bar{p}_{\omega(i-1,n),n}[k]+\alpha[k]\left( h_{i,n}(\bs{d}_i[k])-\frac{c_n}{|\mathcal{I}_n|}\right)\right]^+,~{\rm if}~\trianglelefteq_n {\rm is~}\leq,
\end{cases} \nonumber\\
&\nabla u_i[k]=\nabla_{\bs{x}_i} U_i(\bs{d}_i[k])\label{Update3},
\end{align}
where $\bar{p}_{\omega(i-1,n),n}[k]$ is defined in \eqref{mostrecent}, 
and $\alpha[k]$ is the diminishing step size, given by $\alpha[k]=(1+\beta)/(k+\beta)$ for some non-negative constant $\beta$.

We will soon show that the DyDeNUM Mechanism and the updates in \eqref{Update1}-\eqref{Update3} converge to an optimal solution to the DeNUM Problem and every agent following such updates and reports is a Nash equilibrium.

\subsection{Convergence and Network Utility Maximization}\label{SecVII-B}

In this subsection, we study the properties of the DyDeNUM mechanism. 
\begin{proposition}[Convergence]\label{TT4}
	When Assumptions \ref{Assum2}-\ref{Assum3} hold, if all agents update their messages $\{\bs{m}_i[k]\}$  according to \eqref{Update1}-\eqref{Update3}, each agent $i$'s demand in \eqref{Update1} converges to the optimal action to the DeNUM Problem, i.e., $\lim_{K\rightarrow \infty}\bs{d}_i[K]\rightarrow \bs{x}_i^o$.
\end{proposition}
\begin{IEEEproof}
	Please refer to Appendix \ref{ProofP4}.
\end{IEEEproof}
The convergence proof of Proposition \ref{TT4} is similar to Proposition \ref{T5}. Specifically, the updates according to \eqref{Update1}-\eqref{Update3} together with the DyDeNUM Mechanism essentially constitute an incremental subgradient method \cite{Incremental}, similar to the DeNUM Algorithm.

Different from the DeNUM Algorithm  that distributively solves the dual problem with the indirect decomposition, the DyDeNUM Mechanism with  \eqref{Update1}-\eqref{Update3} exploits a \textit{direct decomposition structure} \cite{NUMTut}.\footnote{
	We present the detailed formulation and analysis in Appendix \ref{Dire}. } In other words, the DyDeNUM Mechanism does not require auxiliary constraints associated with each agent's influence functions $\{h_{i,n}(\bs{x}_i)\}_{n\in\mathcal{N}_i}$ as in the DeNUM Mechanism and R-DeNUM Problem in \eqref{R-NUM}. This eliminates the necessity of Assumption \ref{Assum1}, which is essential for imposing auxiliary constraints.
We are ready to present the following result.

%
%

\begin{theorem}[Nash Equilibrium]\label{TT5}
 When  Assumptions \ref{Assum2}-\ref{Assum3} hold, every agent chooses to update and report according to \eqref{Update1}-\eqref{Update3} is a Nash equilibrium.
\end{theorem}

\begin{IEEEproof}
	Please refer to Appendix \ref{ProofT4}.
\end{IEEEproof}

\rev{Theorem \ref{TT5} implies that the updating and reporting according to \eqref{Update1}-\eqref{Update3} is each agent $i$'s optimal strategy, when all other agents select to do so. Intuitively, supposing that all agents are reporting and updating by \eqref{Update1}-\eqref{Update3}, each agent $i$'s tax is
\begin{align}
\hspace{-0.2cm}\lim_{k\rightarrow \infty}\Pi_{i}[k]
&= \Pi_{i}[0]-\sum_{t=0}^\infty\sum_{j\neq i}\nabla_{\bs x_j} U_j(\bs{d}_j[k])^T(\bs{d}_{j}[k+1]-\bs{d}_{j}[k])\nonumber\\
&\approxeq \Pi_{i}[0]-\sum_{j\neq i} \int_{0}^{\infty}\nabla_{\bs x_j} U_j(\cdot)^Td\bs{d}_{j}(k)\nonumber\\
&=\Pi_{i}[0]-\sum_{j\neq i}  U_j(\bs{d}_j[\infty])+\sum_{j\neq i}  U_j(\bs{d}_j[0]),\label{Approx}
\end{align}
where the approximation in the second line of  \eqref{Approx} is controlled by the step size $\alpha[k]$ in \eqref{Update2}. That is, when the initial step size $\alpha[0]$ is small enough, so is the difference between $\bs{d}_{j}[k+1]$ and $\bs{d}_{j}[k]$. This validates the approximation. }

	\begin{figure}[t]
	\centering
		\includegraphics[scale=0.32]{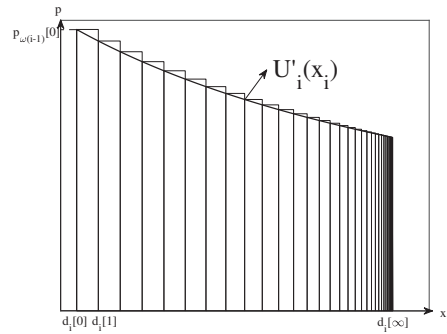}
		\vspace{-10pt}	
		\caption{An illustration of the approximation in Eq. \eqref{Approx}. The sum area of the rectangles $\sum_{t=0}^\infty U_i'({d}_i[k])({d}_{i}[k+1]-{d}_{i}[k])$ approximates the integral $\int_{0}^\infty U'_i(\cdot)d d_i(k)=U_i(d_i[\infty])-U_i(d_i[0])$. }
		\label{DyDe}
	\vspace{-20pt}
\end{figure}

We demonstrate the approximation in Fig. \ref{DyDe}, where the step size $\alpha[k]$ is $350/(k+2000)$. As shown in Fig. \ref{DyDe}, the sum area of the rectangles $\sum_{t=0}^\infty U_i'({d}_i[k])({d}_{i}[k+1]-{d}_{i}[k])$ is a good approximation to the integral $\int_{0}^\infty U'_i(\cdot)d d_i(k)=U_i(d_i[\infty])-U_i(d_i[0])$, since
the relative error is less than $3\%$.\footnote{The relative error is defined as $|U_i(d_i[\infty])-U_i(d_i[0])-\sum_{t=0}^\infty U_i'({d}_i[k])({d}_{i}[k+1]-{d}_{i}[k])|/(U_i(d_i[\infty])-U_i(d_i[0]))$.} The algorithm converges in 31 iterations.
It is possible to further adjust the step size $\alpha[k]$ to make tradeoffs between the error and the convergence speed.

Each agent chooses messages $\{\bs{m}_i[k]\}_{k\in\mathbb{N}}$ to maximize her convergent payoff, given by $U_i(\bs{d}_i[\infty])-\Pi_{i}[\infty]$. 
From \eqref{Approx}, since $\Pi_i[0]$ and $\sum_{j\neq i}U_j(\bs{d}_j[0])$ are constant, each agent chooses to maximize her utility $U_i(\bs{d}_i[\infty])$ plus the term $\sum_{j\neq i}U_j(\bs{d}_j[\infty])$, which equals the network utility. Moreover,  from Proposition \ref{TT4}, if all agents follow the updates in \eqref{Update1}-\eqref{Update3}, the DyDeNUM Mechanism achieves the maximal network utility. Therefore, when all agents other than $i$ update and report  according to \eqref{Update1}-\eqref{Update3}, it is always agent $i$'s optimal strategy to do so.

\subsection{Computation of the Initial Tax and Individual Rationality}
In this subsection, we discuss that it is possible to further design $\{\Pi_i[0]\}_{i\in\mathcal{I}}$ in a way to achieve the individual rationality (E2). Specifically, we can consider a distributed algorithm similar to the updates in \eqref{Update1}-\eqref{Update3}. Such an algorithm computes the maximal network utility of the system excluding agent $i$.\footnote{Due to the space limit, we present the details of such an algorithm in Appendix \ref{Algo2}.} The resulted initial tax $\Pi_i[0]$ together with \eqref{Approx} makes each agent $i$ receive a VCG-type tax, and can thus achieve the individual rationality as follows. 
\begin{proposition}\label{P5}
When Assumptions \ref{Assum2}-\ref{Assum5} hold, the DyDeNUM Mechanism satisfies the individual rationality (E2).
\end{proposition}
\begin{IEEEproof}
	Please refer to Appendix \ref{ProofP5}.
\end{IEEEproof}
However, our DyDeNUM Mechanism cannot guarantee the budget balance, which is one of the disadvantages of the VCG-type taxation. In a nutshell, the DyDeNUM Mechanism exploits the direct composition structure to avoid the necessity of Assumption \ref{Assum1}, however, at the cost of the budget balance.

\section{Application: Fog-Based User-Provided Network}\label{UPN}

\begin{figure}[t]
	\centering
		\includegraphics[scale=0.4]{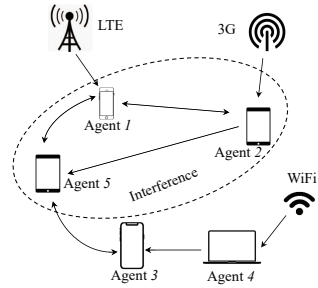}
		\vspace{-10pt}
		\caption{Examples of agent interactions in the fog-based UPNs. Each agent can concurrently consume data over multiple and multi-hop paths, and serve as a relay/gateway for other  agents.  }
		\label{example}
\end{figure}

In this section, we consider the fog-based user-provided network (UPN) as a concrete example of the DeNUM framework, to demonstrate the effectiveness of the DeNUM Mechanism, the DeNUM Algorithm, and the DyDeNUM Mechanism. Such an application is motivated by the Open Garden framework \cite{Open,Meng}. As shown in Fig. \ref{example}, in a UPN, near-by agents (who are mobile users) can form a mesh network through Bluetooth or Wi-Fi Direct, and share their Internet access capabilities among each other. 

UPNs can exploit diverse network resources and thus improve the overall network performance.
However, the success of such services relies on an appropriate economic mechanism that can provide incentives for providing services and cope with information asymmetries.\footnote{Only one existing work considers an incentive mechanism for achieving the optimum of the corresponding DeNUM Problem, but 
	assumes the complete information (of both utility and constraint information) \cite{Open}. Note that even without considering information asymmetries of constraints, existing mechanisms (e.g. \cite{VCGKelly,scalar,Jain2010,Multicast,power,Surrogate,learning,PEV,sharma2012local,General}) are not applicable to such an application. Specifically, such an application consists of more complicated constraints than those in \cite{VCGKelly,scalar,Jain2010,Multicast,power,Surrogate,learning,PEV,sharma2012local}, whereas this is no convergent algorithm in  \cite{General}.}

\subsection{System Model}
\textit{Wireless mesh network \cite{Open}.} Consider a mesh network that is described by a directed graph $G=(\mathcal{I}, \mathcal{E})$, where $\mathcal{I}$ denotes the set of users and  $\mathcal{E}$ denotes the set of communication links. We define $\mathcal{B}_{(i\leftarrow j)}$ as the set of all direct links that will interfere with link $(i\leftarrow j)$. Let $C_{ij}$ be the capacity of link $(i\leftarrow j)$.

\textit{User decisions.} Let $x_{i\leftarrow j}(n)$ denote the amount of data user $j$'s to her one-hop downstream neighbor $i$, where $(n)$ represents that such a unicast session originates from the Internet (e.g., a web/content server) and will end at user $n\in\mathcal{I}$. Let $\bs{x}_i^{\rm In}=\{x_{i\leftarrow j}(n)\}_{j\in\mathcal{I}^{\rm In}(i), n\in\mathcal{I}}$ be user $i$'s upstream data vector and let $\bs{x}_i^{\rm Out}=\{x_{j\leftarrow i}(n)\}_{j\in\mathcal{I}^{\rm Out}(i), n\in\mathcal{I}}$ denote user $i$'s downstream data vector, where $\mathcal{I}^{\rm In}(i)$ and $\mathcal{I}^{\rm Out}(i)$ are sets of user $i$'s upstream and downstream one-hop neighbors, respectively. Let $y_i(n)$ be the amount of data user's $i$ downloaded from the Internet for user $n$. Based on the protocol
interference model  \cite{Open}, we assume that a transmission
over link $(i\leftarrow j)\in \mathcal{E}$ is successful only if all other links $(k\leftarrow m)$ in $\mathcal{B}_{(i\leftarrow j)}$ are idle. That is, user traffic decisions need to satisfy
\begin{align}
&\sum_{n\in\mathcal{I}}\frac{x_{i\leftarrow j}(n)}{C_{ij}}+\sum_{(k\leftarrow m)\in\mathcal{B}_{(i\leftarrow j)}}\sum_{n\in\mathcal{I}}\frac{x_{k\leftarrow m}(n)}{C_{km}}\leq 1, \nonumber\\
&~~~~~~~~~~~~~~~~~~~~~~~~~~~~~~~~~~~~~~~~~ \forall (i\leftarrow j)\in\mathcal{E}.\label{Protocol}
\end{align}
This means that in each time period (the length of which is normalized to 1), the total amount of transmission time of link $(i\leftarrow j)$ and all links in $(k\leftarrow m)\in\mathcal{B}_{(i\leftarrow j)}$ cannot exceed $1$. In our DeNUM framework, constraint \eqref{Protocol} belongs to a system constraint.

\textit{User utility and cost.} Each user $i$ has an increasing and strictly concave utility ${U}_i(r_i)$,
where $r_i$ is the user $i$'s received data, given by
$r_i(\bs{x}_i^{\rm In},y_i(i))=\sum_{j\in\mathcal{I}^{\rm In}(i)}x_{i\leftarrow j}(i)+y_i(i)$.
Each user has an increasing strictly convex cost function $C_i(\bs{y}_i,\bs{x}_i^{\rm Out})$, which captures the energy consumption and payments of the mobile date services.

 Specifically,
each user $i\in\mathcal{I}$ has a maximum energy budget $E_i$. Let $e_{j\leftarrow i}^s$ be the energy that user $i$ consumes when she sends one byte to $j$. Let $e_{i\leftarrow j}^r$ be the energy that user $i$ consumes when she receives one byte from user $j$. Finally, $e_{0i}$ is an energy consumption when node $i$ downloads one byte from the Internet.
The total consumed energy $e_i$ for each user $i$ thus is:
\begin{align}
e_i=&\sum_{j\in \mathcal{I}^{\rm In}(i)}e_{i\leftarrow j}^r\sum_{n\in\mathcal{I}}x_{i\leftarrow j}(n)+\sum_{j\in\mathcal{I}^{\rm Out}(i)}{e}_{j\leftarrow i}^s\sum_{n\in\mathcal{I}}x_{j\leftarrow i}(n)\nonumber\\
&~~~~+e_{0i}\sum_{n\in\mathcal{I}}y_i{(n)}.
\end{align} 
The energy cost function is given by\cite{Open}
\begin{align}
C_i(e_i)=\frac{\delta_i}{E_i-e_i},
\end{align}
where $\delta_i$ is a normalization parameter indicating user $i$'s sensitivity in energy consumption.

 We define each user $i$'s payoff as, 
\begin{align}
\hspace{-0.42cm}J_i(r_i,\bs{y}_i,\bs{x}_i^{\rm Out},\bs{x}_i^{\rm In})\triangleq {U}_i(r_i(\bs{x}_i^{\rm In},y_i(i)))-C_i(\bs{y}_i,\bs{x}_i^{\rm Out}),
\end{align}
which is strictly concave.

\subsection{Problem Formulation}

Note that users' utilities are coupled through their decision variables $\bs{x}=\{x_{i\leftarrow j}(n)\}_{(i\leftarrow j)\in\mathcal{E},n\in\mathcal{I}}$. 
Hence, we further introduce auxiliary variables $\bs{x}^r=\{x_{i\leftarrow j}^r(n)\}_{(i\leftarrow j)\in\mathcal{E},n\in\mathcal{I}}$ and $\bs{x}^s=\{x_{i\leftarrow j}^s(n)\}_{(i\leftarrow j)\in\mathcal{E},n\in\mathcal{I}}$, where $x_{i\leftarrow j}^s(n)$ is agent $j$'s decision variable (of sending data) and $x_{i\leftarrow j}^r(n)$ is user $i$'s decision variable (of receiving data). 
Let $\bs{x}_i^{r, \rm In}=\{x^r_{i\leftarrow j}(n)\}_{j\in\mathcal{I}^{\rm In}(i), n\in\mathcal{I}}$ be user $i$'s upstream data vector and let $\bs{x}_i^{s, \rm Out}=\{x^s_{j\leftarrow i}(n)\}_{j\in\mathcal{I}^{\rm Out}(i), n\in\mathcal{I}}$ denote user $i$'s downstream data vector.

We formulate the DeNUM problem as
\begin{subequations}
	\begin{align}
	&\max_{\bs{r},\bs{x}^s,\bs{x}^r,\bs{y}}~\sum_{i\in\mathcal{I}}\left[{U}_i(r_i(\bs{x}_i^{r, \rm In},y_i(i)))-C_i(\bs{y}_i,\bs{x}_i^{s, \rm Out})\right]\label{A1}\\
	&~{\rm s.t.}~~~r_i=\sum_{j\in\mathcal{I}^{\rm In}(i)}x_{i\leftarrow j}^r(i)+y_i(i),~\forall i\in\mathcal{I},\label{A2}\\
	&~~~~~~\sum_{n\in\mathcal{I}}y_{i}(n)\leq C_{0i},~\forall i\in\mathcal{I},\label{A3}\\
	&~~~~~~\sum_{n\in\mathcal{I}}\left[\frac{x_{i\leftarrow j}^s(n)}{C_{ij}}+\sum_{(k\leftarrow m)\in\mathcal{B}_{(i\leftarrow j)}}\!\!\!\!\frac{x_{k\leftarrow m}^s(n)}{C_{km}}\right]\leq 1,\nonumber\\
	&~~~~~~~~~~~~~~~~~~~~~~~~~~~~~~~~~~~~~~~~~~~~~\forall (i\leftarrow j)\in\mathcal{E},\label{A4}\\
	&~~~~~~y_i(n)+\sum_{j\in\mathcal{I}^{\rm In}(i)}x^r_{i\leftarrow j}(n)=\sum_{j\in\mathcal{I}^{\rm Out}(i)}\!x^s_{j\leftarrow i}(n),\nonumber\\&~~~~~~~~~~~~~~~~~~~~~~~~~~~~~~~~~~~~~~~~~~\forall i, n\in\mathcal{I}, n\neq i,\label{A6}\\
	&~~~~~~x_{i\leftarrow j}^s(n)=x_{i\leftarrow j}^r(n),~\forall (i\leftarrow j)\in\mathcal{E}.\label{A7}
	\end{align}
\end{subequations}
Constraint \eqref{A3} characterizes agent $i$'s downlink capacity constraint with $C_{0i}$ being user $i$'s downlink capacity.
Constraint \eqref{A7} is the consistency constraint that equalizes the decision variables of user $i$ and user $j$.

In such a formulation, the constraints in \eqref{A2}-\eqref{A3} and \eqref{A6} are individual constraints whereas the constraints in \eqref{A4} and \eqref{A7} are system constraints.

\subsection{Numerical Study}

%
%
%
\subsubsection{Simulation Setup}
In this subsection, we consider a setting with 5 agents where agent 2 does not have an Internet connection. Every user $i\in\mathcal{I}$ has an  $\alpha$-fair utility function ${U}_i(r_i)=r_i^{1-\alpha_i}/(1-\alpha_i)$, which is widely used in the literature \cite{Ref1}. Parameters $\alpha_i\in[0,1]$ represent the willingness to pay of the different utilities. 

User energy consumption model is based on empirical data. Specifically, we consider a set of $|\mathcal{I}|=5$ users, randomly placed in a geographic area, and study their interactions for a time period of $T=120$ seconds. 
We assume that users communicate with each other using WiFi Direct. The achievable capacity between two users decreases with their distance $d_{ij}$. We set the WiFi Direct capacity to be $C_{ij}=100\log(1+0.9/d_{ij}^2)$. Users are placed in the $[0,30m]\times[0,30m]$ plane randomly, with a uniform distribution.

For mobile devices, the energy consumed by a data transfer is proportional to the size of the data and the transmission power level.  Typically, the energy consumption (per MByte) of WiFi transmissions is smaller compared to LTE transmissions, which, in turn, is smaller than 3G transmissions. We consider here an average energy consumption of $e_{0i}=100$ Joules/MByte when user $i$ has a 3G Internet access connection, $e_{0i}=100$ Joules/MByte, 
$e_{0i}=4.65$ Joules/MByte for an LTE connection, and $e_{0i}=2.85$ Joules/MByte for a WiFi connection \cite{Ref2,Ref3,Ref4}. For WiFi direct links, we assume that the energy consumption per MByte increases with the distance (since the achievable rate decreases with the distance), in the form of $e_{i\leftarrow j}^s=(2.77+0.008d_{ij})$ Joules/MByte.

We consider a setting where user $1$ has a LTE connection, user $2$ does not have connection, users $3$ and $4$ have 3G connections, and user $5$ have WiFi connections. Specifically, the downlink capacities for all agents are $C_{0i}=\{12.7, 0.0, 1.0, 1.0, 4.12\}$Mbps, respectively, which are based on the field experiments in \cite{Ref5,Ref6,Ref7}.

\begin{figure}[t]
		\subfigure[]{		\includegraphics[scale=.21]{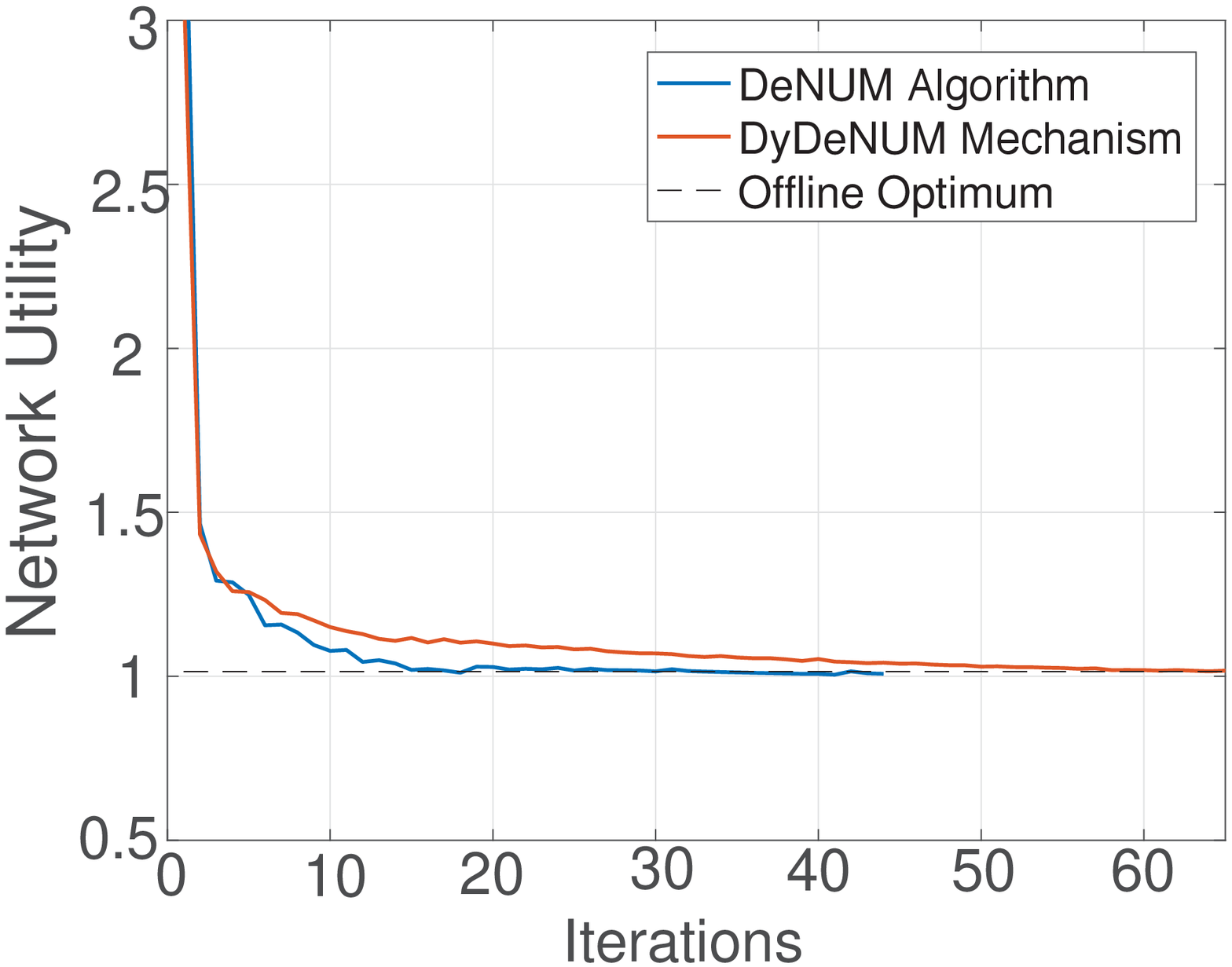}}
		\subfigure[]{	\includegraphics[scale=.21]{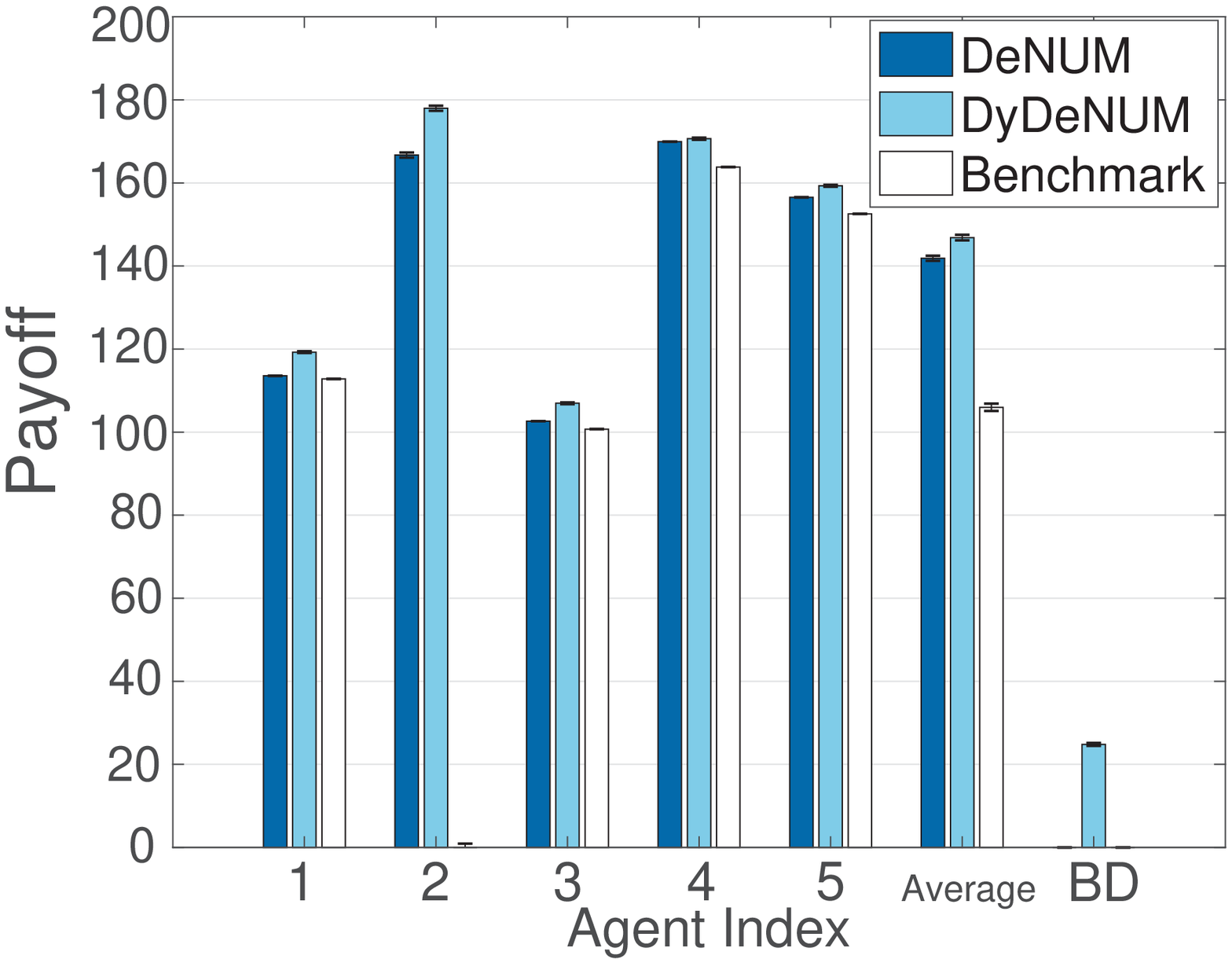}}
		\vspace{-8pt}
		\caption{Simulation results of a UPN with 5 agents: (a) evolution of the network utility produced by the DeNUM Algorithm and the DyDeNUM Mechanism; (b) payoffs and budget deficiency (`BD') due to compensations  achieved/incurred by the DeNUM Mechanism, the DyDeNUM Mechanism, and a benchmark solution.}
		\label{simulation}
	\vspace{-15pt}
\end{figure}

\subsubsection{Simulation Results}
In Fig. \ref{simulation} (a), we plot agents' network utility  achieved by the DeNUM Algorithm and the DyDeNUM Mechanism
in each iteration. We see that the DeNUM Mechanism converges 
to the offline optimum within 20 iterations and the DyDeNUM Mechanism converges within 60 iterations. The DyDeNUM Mechanism converges more slowly since it requires a small step size in \eqref{Update2} to achieve a reasonable approximation in \eqref{Approx}.
In addition, before the convergence, the network utility is
even larger than the respective optimal value, which is because the produced aggregate payoff is not feasible until convergence. 

Second,
in Fig. \ref{simulation} (b), we consider a benchmark scheme in which each agent can only access her own cellular downlink (which is equivalent to not participating in any mechanism and bear a constraint in \eqref{Out}). We study the performances of the DeNUM Mechanism and the DyDeNUM Mechanism, compared with the benchmark.\footnote{The results represent the average obtained
 over 100 experiments for different user locations and hence
 user distances.}
 We observe that, for each agent $i$,
  both mechanisms always
 improve upon the benchmark payoff, which implies that both mechanisms achieve the individual rationality (E2). In addition, the DeNUM Mechanism improves the average payoff by $34\%$. Although both the DyDeNUM Mechanism and the DeNUM Mechanism achieve the same maximal network utility, the DyDeNUM Mechanism achieves a higher payoff for each agent. This is because the system designer needs to compensate each agent to participate in the DyDeNUM Mechanism and therefore incurs a budget deficiency, as also shown in Fig. \ref{simulation} (b).








\section{Conclusions}\label{Conclusion}

In this paper, we proposed a new economic mechanism framework for solving the decomposable NUM Problems in network sharing. Our proposed DeNUM Mechanism can cope with
 agents' strategic behaviors and private utility and constraint information,  with the desirable economic
properties including efficiency,
individual rationality, and budget balance.
 In addition, we
 proposed a distributed low-complexity DeNUM Algorithm provably convergent to the equilibrium of the DeNUM Game. 
 We further designed a DyDeNUM Mechanism that achieve the network utility maxima even if the monitorable influence assumption is not satisfied but at the cost of the balanced budget. There are several directions for extensions. Since our mechanisms are susceptible to collusive agents, one may ask how to design \textit{group-strategyproof} mechanisms for the NUM framework.



\appendices

\section{Model Extensions}

\subsection{Decoupling of Coupled Utilities}\label{DCU}

To decouple the coupled agents' utilities, we can reformulate the problem by introducing auxiliary variables and auxiliary consistency equality constraints. We consider the following illustrative example:
\begin{example}
Suppose there are two agents having the coupled utilities $U_1(\bs{x}_1,\bs{y})$ and $U_2(\bs{x}_2,\bs{y})$, respectively. The DeNUM Problem is
\begin{align}
\max_{\bs{x}_1,\bs{x}_2,\bs{y}}U_1(\bs{x}_1,\bs{y})+U_2(\bs{x}_2,\bs{y})\label{coupled}.
\end{align}

To decouple the objectives, we can introduce the auxiliary variables $\bs{y}_1$ and $\bs{y}_2$ and an additional equality system constraint. We hence have the following equivalent reformulation:
\begin{subequations}
\begin{align}
&\max_{\bs{x}_1,\bs{y}_1,\bs{x}_2,\bs{y}_2}~U_1(\bs{x}_1,\bs{y}_1)+U_2(\bs{x}_2,\bs{y}_2)\\
&~~~~~{\rm s.t.}~~~~~\bs{y}_1=\bs{y}_2.
\end{align}
\end{subequations}
The reformulation is not only equivalent to \eqref{coupled} but also can be captured by the decoupled formulation of the DeNUM Problem in \eqref{NUM}.
\end{example}

\subsection{Decoupling of Coupled System Constraints}\label{DCC}

To decouple the coupled agents' system constraints, we can reformulate the problem by introducing auxiliary variables and auxiliary consistency equality constraints, similar as in Appendix \ref{DCU}. We consider the following illustrative example:
	\begin{example}
		Suppose there are two agents having the coupled utilities $U_1(\bs{x}_1,\bs{y})$ and $U_2(\bs{x}_2,\bs{y})$, respectively. The DeNUM Problem is
\begin{align}
&\max_{\bs{x}_1,\bs{x}_2}~U_1(\bs{x}_1)+U_2(\bs{x}_2)\\
&~{\rm s.t.}~~{h}(\bs{x}_1,\bs{x}_2)\trianglelefteq c.\label{CC}
\end{align}
		To decouple the system constraint in \eqref{CC}, we can introduce the auxiliary variables $\bs{y}_1$ and $\bs{y}_2$, auxiliary local constraints for two agents, an additional equality system constraint. We hence have the following equivalent reformulation.
		\begin{subequations}
			\begin{align}
			&\max_{\bs{x}_1,\bs{y}_1,\bs{x}_2,\bs{y}_2}~U_1(\bs{x}_1,\bs{y}_1)+U_2(\bs{x}_2,\bs{y}_2)\\
			&~~~~~{\rm s.t.}~~~~~\bs{y}_1=\bs{y}_2,\\
			&~~~~~~~~~~~~~~{h}(\bs{x}_i,\bs{y}_i)\trianglelefteq c,~\forall i\in\{1,2\}.
			\end{align}
		\end{subequations}
		The reformulation is not only equivalent to \eqref{coupled} but also can be captured by the decoupled formulation of the DeNUM Problem in \eqref{NUM}.
	\end{example}
	
\subsection{Private System Constraint Parameters}\label{UC}

To tackle with the private system constraint parameters, we can introduce auxiliary local constraint and auxiliary equality system constraint. We use the following example to illustrate this.
\begin{example}
Consider the following DeNUM Problem:
\begin{subequations}
\begin{align}
&\max_{\bs{x}_1,\bs{x}_2}~U_1(\bs{x}_1)+U_2(\bs{x}_2)\label{privatec}\\
&~~{\rm s.t.}~\bs{x}_1\in\mathcal{X}_1,\\
&~~~~~~~h_{1}(\bs{x}_1)+h_2(\bs{x}_2)\leq c.
\end{align}
\end{subequations}
Suppose parameter $c$ is agent $1$'s private information. Let us introduce an auxiliary variable $y$ and $\hat{\mathcal{X}}_1=\{(\bs{x}_1,y):\bs{x}_1\in\mathcal{X}_1, h_1(\bs{x}_1)+y\leq c\}$. Consider the following reformulated problem:
\begin{subequations}
\begin{align}
&\max_{\bs{x}_1,\bs{x}_2,y}U_1(\bs{x}_1,y)+U_2(\bs{x}_2)\\
&~~{\rm s.t.}~(\bs{x}_1,y)\in\hat{\mathcal{X}}_1\\
&~~~~~~~h_2(\bs{x}_2)-y=0\label{c0}.
\end{align}
\end{subequations}
We see that such reformulation is equivalent to \eqref{privatec} and can be captured by the DeNUM Problem in \eqref{NUM}, since the new system constraint's budget is globally known as $0$.
\end{example}
\subsection{An Example of the Decomposable NUM Problem}\label{ExampleC}

We consider an illustrative network sharing example to show that a practical DeNUM Problem fits in the aforementioned setting:
\begin{example}
Consider a fog computing system with 2 agents and CPU and RAM budget constraints. Agent $1$ is the data center owner and agent $2$ requires a fixed amount of each resource to accomplish two types of jobs. We assume that both types of jobs are divisible. Agent $2$ requires $1$ CPU and $4$ GB of RAM for each unit amount of job $a$ and $3$ CPU and $2$ GB of RAM for each unit amount of job $b$. Agent $2$ has $9$ CPUs and $18$ GB of RAM. The NUM Problem is therefore formulated as
\begin{subequations}
	\begin{align}
	&\max_{\bs{x}_1, \bs{x}_2}~~U_1(\bs{x}_1)-C( \bs{x}_2)\\
	&~~{\rm s.t.}~~~~~x_{1,a}+3x_{1,b}-x_{2,\rm C}=0\label{ex-sys1}\\
	&~~~~~~~~~~4x_{1,a}+2x_{1,b}-x_{2,\rm R}=0\label{ex-sys2}\\
	&~~~~~~\bs{x}_1\succeq0,~x_{2,\rm C}\leq 9,~x_{2,\rm R}\leq 18\label{ex-pric1}
	\end{align}
\end{subequations}
Suppose that agent $1$ provides $x_{1,\rm C}$ CPUs and $x_{1,\rm R}$ GB of RAM;
the accomplished amounts of jobs $a$ and $b$ for agent $2$ are $x_{2,a}$ and $x_{2,b}$. Hence, agent $1$'s local constraint is $\mathcal{X}_1=\{(x_{1,\rm C},x_{1,\rm R}): 0\leq x_{1,\rm C}\leq 9, 0\leq x_{1,\rm R}\leq 18 \}$ which is not known by others;
agent $2$'s influences functions are $h_{2,1}(\bs{x}_2)=x_{2,a}+3x_{2,b}$ and $h_{2,2}(\bs{x}_2)=4x_{2,a}+2x_{2,b}$ for the two system constraints, respectively, indicating that the action of accomplishing jobs consuming the resources. Note that, in this scenario, agent $2$ may not know agent $1$'s resource budget and 
agent $1$ may not know how many resources each job requires (local constraints and influence functions are unknown). But agent $1$ can observe how many resources are consumed afterwards (Assumption \ref{Assum1} is satisfied since the output of the influence functions are monitorable). 
\end{example}

\section{Proofs}

\subsection{Proof of Proposition \ref{T1}}\label{ProofP1}

  To prove Proposition \ref{T1}, we first prove the following lemma:
\begin{lemma}\label{L1}
	If the maximal achievable network utility is less than $\sum_{i\in\mathcal{I}}U_i(\bs{0})$, no mechanism can yield (E2) and (E3).
\end{lemma}

The intuition is that the constraint in \eqref{Out} ensures each agent can achieves at least a $\sum_{i\in\mathcal{I}}U_i(\bs{0})$ after opting out of the mechanism. If the maximal achievable network utility is lower than $\sum_{i\in\mathcal{I}}U_i(\bs{0})$, no money injection (budget balance) leads to that circumstance where at least one agent is worse off than receiving $U_i(\bs{0})$. Note that this result applies to all (game-theoretic) solution concepts, not limited to the Nash equilibrium or the GNE.
\begin{IEEEproof}
Let $(\bs{x}^*, \bs{\Pi}^*)$ be an equilibrium (not necessarily an Nash equilibrium or a GNE) actions and payments of a mechanism. In order to achieve the individual rationality, we must have
\begin{align}
U_i(\bs{x}_i^*)-\Pi_i^*\geq \max_{\bs{x}_i\in\mathcal{X}_i^{\rm Out}} U_i(\bs{x}_i). \label{N}
\end{align}
By the definition of $\mathcal{X}_i^{\rm Out}$ in \eqref{Out} and the fact that $h_{i,n}(\bs{x}_i)=U_i(\bs{x}_i)=0,$ for all $(i,n)\in\{(i,n):i\in\mathcal{I}_n\}$, it follows that
\begin{align}
U_i(\bs{x}_i^*)-\Pi_i^*\geq \max_{\bs{x}_i\in\mathcal{X}_i^{\rm Out}} U_i(\bs{x}_i)\geq U_i(\bs{0}).
\end{align}
To achieve the (weak) budget balance (which is a ), we have $
\sum_{i\in\mathcal{I}}\Pi_i^*= 0$.
Hence, we must have
\begin{align}
\sum_{i\in\mathcal{I}}U_i(\bs{x}^*_i)-\sum_{i\in\mathcal{I}}\Pi_i^*&\geq \sum_{i\in\mathcal{I}}U_i(\bs{x}^*_i)\geq \sum_{i\in\mathcal{I}}U_i(\bs{0}),
\end{align}
which implies that weak budget balance, individual rationality, and social optimum at any equilibrium cannot be satisfies when the maximal achievable network utility (social welfare) is negative, no matter what mechanism rules are.
\end{IEEEproof}

We then prove the proposition  by construction, showing that (E2) and (E3) cannot be satisfied in two examples:
\begin{enumerate}
\item Consider a DeNUM Problem:
	\begin{align}
	&\max_{\bs{x}}~\sum_{i\in\mathcal{I}}\log(1+x_i)~~{\rm s.t.}~\sum_{i\in\mathcal{I}}x_i~\trianglelefteq~ c.
	\end{align}
	The optimal solution is $x_i^*=c/I$ for each $i\in\mathcal{I}$
	regardless of the fact that the constraint is equality or inequality.
	The maximal network utility is $I\log(1+c/I)$, which is negative (less than $\sum_{i\in\mathcal{I}}U_i(\bs{0})=0$) if and only if $c$ is negative. Therefore, 
	 according to Lemma \ref{L1}, in this example, no mechanism that can yield (E2) and (E3).

\item  Consider an another NUM problem:
\begin{subequations}
	\begin{align}
	&\max_{x_1, x_2}~\log(1+x_1)+\log(1+x_2),\\
	&~{\rm s.t.}~~\begin{bmatrix}
	1 & 1 \\
	1 & 0.5 
	\end{bmatrix}\begin{bmatrix}
	x_1 \\
	x_2 
	\end{bmatrix}=\begin{bmatrix}
	0 \\
	0.25 
	\end{bmatrix}.
	\end{align}
\end{subequations}
		There is only one feasible (and hence the optimal) solution $(x_1^*=0.5, x_2^*=-0.5)$,
	which leads to a negative network utility (less than $\sum_{i\in\mathcal{I}}U_i(\bs{0})=0$). Hence, by Lemma \ref{L1}, in this instance, no mechanism that can yield (E2) and (E3).
	\end{enumerate}

To sum up, the first example shows that when $c$ is negative, it is possible that the properties in (E2) and (E3) cannot be satisfied at the same time for both equality and inequality constraints. The second example shows that, in an equality constraint case, no mechanism that can yield (E2) and (E3). Combining the results of the two cases, we complete the proof.

\subsection{Proof of Theorem \ref{T3}}\label{ProofT1}
Let $f_{\mathcal{X}_i}(\bs{x}_i)$ denote a convex and continuously differentiable function that characterizes the set $\mathcal{X}_i$ as, $\bs{x}_i\in\mathcal{X}_i$ if and only if $f_{\mathcal{X}_i}(\bs{x}_i)\leq 0$.\footnote{By \cite{boyd2004convex}, there always exists such a function $f_{\mathcal{X}_i}(\bs{x}_i)$ for any convex set $\mathcal{X}_i$.}
Since the DeNUM Problem is convex and satisfies the Slater's condition,  the DeNUM Problem's sufficient and necessary KKT conditions for optimality are, for any $(i,n)\in\{(i,n): i\in\mathcal{I}_n\}$,
\begin{subequations}\label{KKTNUM}
	\begin{align}
	\hspace{-0.5cm}\nabla_{\bs{x}_i}U_i(\bs{x}_i)-\!\!\sum_{n\in\mathcal{N}_i}\!\!\lambda_{i,n}\nabla_{\bs{x}_i} h_{i,n}(\bs{x}_i) -\gamma_i \nabla_{\bs{x}_i} f_{\mathcal{X}_i}(\bs{x}_i)&=\bs{0},\label{KKT1}\\
	\Xi_n\left(\lambda_{n}\right)\left(\sum_{i\in\mathcal{I}}h_{i,n}(\bs{x}_i)-c_{n}\right)&=0,\label{KKT2}\\
	\sum_{i\in\mathcal{I}}h_{i,n}(\bs{x}_i)&\trianglelefteq_n\!\! c_{n},\label{KKT3}\\
	\gamma_if_{\mathcal{X}_i}(\bs{x_i})&= 0,\label{KKT4}\\
	\Xi_n(\lambda_{n}), \gamma_i&\geq 0,\label{KKT5}\\
	f_{\mathcal{X}_i}(\bs{x}_i)&\leq 0,\label{KKT6}
	\end{align}
\end{subequations}
where
\begin{align}
\Xi_n(a)\triangleq\begin{cases}
a,~~{\rm if}~\trianglelefteq_n~{\rm is}~\leq,\\
1,~~{\rm if}~\trianglelefteq_n~{\rm is}~=.
\end{cases}
\end{align}
Let  $(\bs{x}^o,\bs\lambda^o,\bs{\gamma}^o)$ denote the solution to the  KKT conditions in \eqref{KKTNUM}.

On the other hand,  we reformulate agent $i$'s APM Problem into the following equivalent form:
\begin{subequations}\label{APM1}
\begin{align}
&\max_{\bs{x}_i\in\mathcal{X}_i,\bs\tau_i,\bs t_i}~U_i(\bs{x}_i)-\sum_{n\in\mathcal{N}_i}p_n^{*}t_{i,n}\\
	&~~{\rm s.t.}~~~h_{i,n}(\bs{x}_i)\trianglelefteq_n t_{i,n},~\forall n\in\mathcal{N}_i\label{C11},\\
	&~~~~~~~~~t_{i,n}=\tau_{i,n}-\frac{\sum_{j\in\mathcal{I}_n} \tau_{j,n}-c_n}{|\mathcal{I}_n|},~\forall n\in\mathcal{N}_i.\label{C12}
\end{align}
\end{subequations}
More specifically, we assume that $\bs t_{i}$ is also agent $i$'s decision variable and introduce the constraint in \eqref{C12}.
It is readily verified that the Problem in \eqref{APM1} is convex and the corresponding Slater's conditions  are also satisfied (by Assumption \ref{Assum2}). Therefore, 
the sufficient and necessary KKT conditions for each agent $i$'s APM Problem are, for each $ n\in\mathcal{N}_i$,
        \begin{subequations}\label{KKTAPM}
		\begin{align}
		  \hspace{-0.4cm}\nabla_{\bs{x}_i}U_i(\bs{x}_i)-\!\!\sum_{n\in\mathcal{N}_i}\!\!\lambda_{i,n}\nabla_{\bs{x}_i} h_{i,n}(\bs{x}_i) -\gamma_i \nabla_{\bs{x}_i} f_{\mathcal{X}_i}(\bs{x}_i)&=\bs{0},\label{KKT111}\\
		\lambda_{i,n}-p_n^{*}&=0,\\
		\Xi_n\left(\lambda_{i,n}\right)(h_{i,n}(\bs{x}_i)-t_{i,n})&=0, 
	\label{KKT11}\\
		h_{i,n}(\bs{x}_i)&\trianglelefteq_n t_{i,n},\label{KKT12}\\
		\gamma_if_{\mathcal{X}_i}(\bs{x}_i)&= 0, \label{KKT112}\\
		\Xi_n(\lambda_{i,n}), \gamma_i&\geq 0, \label{KKT122}\\
		f_{\mathcal{X}_i}(\bs{x}_i)&\leq 0,
		\end{align}
		\end{subequations}
		where $\lambda_{i,n}$ is the dual variables corresponding to constraints in \eqref{C11}.
Agents' GNE decisions for described by $(\bs{x}^*,\{\lambda_{i,n}\}_{i,n}^*,\bs{t}^*,\bs{\gamma}^*)$ satisfy \eqref{KKTAPM} and that $\sum_{i\in\mathcal{I}_n}t_{i,n}^*=c_n$ due to \eqref{constraint1}. We are ready to prove the existence and efficiency of the GNEs.
\subsubsection{Existence}

Assumption \ref{Assum2} ensures that there exists an optimal solution to the DeNUM Problem. For any $(\bs{x}^o,\bs{\lambda}^o,\bs{\gamma}^o)$  to the  KKT conditions in \eqref{KKTNUM}, we will show that the strategy profile $(\bs{x}^*, \bs{m}^*)$ such that, for all $(i,n)\in\{(i,n): i\in\mathcal{I}_n\}$,
\begin{align}
\bs{x}_i^*=\bs{x}_i^o,~\tau_{i,n}^*=h_{i,n}(\bs{x}_i^o),~p_{i,n}^*=\lambda_n^o, \label{NEe}
\end{align}
is a GNE of the DeNUM Game. First, it is easy to see that \eqref{NEe} and \eqref{KKTNUM} assure \eqref{KKT111}, \eqref{KKT112}, and \eqref{KKT122}. In addition, let $t_{i,n}$ satisfy \eqref{constraint1}. Then, we see that $h_{i,n}(\bs{x}_i^o)=t_{i,n}$ if $\sum_{i}h_{i,n}(\bs{x}_i^o)=c_n$ and $h_{i,n}(\bs{x}_i^o)<t_{i,n}$ if $\sum_{i}h_{i,n}(\bs{x}_i^o)<c_n$, which satisfies the conditions in \eqref{KKT11} and \eqref{KKT12}. Therefore, there exists at least one GNE.\footnote{There are multiple existent GNEs in general, mainly resulting from the possibility of multiple $\bs{x}^o$ and $\bs{\lambda}^o$.}

\subsubsection{Efficiency}
		
		It is readily verified that, $(\bs{x}^*,\{\lambda_n=p_n^*=\lambda_{i,n}^*\}, \bs{\gamma}^*)$ also satisfies the conditions in \eqref{KKT1}, \eqref{KKT4}, and \eqref{KKT5}. In addition, $\sum_{i\in\mathcal{I}_n}t_{i,n}^*=c_n$ together with \eqref{KKT11} and \eqref{KKT12} further indicates that $(\bs{x}^*,\{\lambda_n=p_n^*=\lambda_{i,n}^*\}, \bs{\gamma}^*)$ also satisfies \eqref{KKT2} and \eqref{KKT3}. Therefore, any GNE leads to the optimal solution.

\subsection{Proof of Theorem \ref{T4}}\label{ProofT2}

The main idea of the proof of Theorem \ref{T4} is to show that, regardless of other agents' strategies, there always exists a strategy $(\bs{x}_i,\bs{m}_i)$ for each agent $i$ that yields exactly the same payoff as no participating the mechanism.

We define
$V_i(\bs{m})\triangleq \max_{ \bs{x}_i\in\tilde{\mathcal{X}}_i(\bs{m}_i;\bs{m}_{-i})} J_i(\bs{x}_i,\bs{m}_i)$.
From \eqref{NEprice}, at a GNE, agent $i$'s payoff can be rewritten as
	\begin{align}
	V_i(\bs{m}^*)&\triangleq\!\!\max_{\substack{\bs{m}_i\\ \bs{x}_i\in\tilde{\mathcal{X}}_i(\bs{m}_i;\bs{m}_{-i}^*)}} J_i(\bs{x}_i,\bs{m}^*).
	\end{align}
	At a GNE, agent $i$ can always submit her message $\hat{\bs{m}}_i=(\hat{\bs\tau}_i,{\bs p}_i^*)$ where $\hat{\tau}_{i,n}=\frac{\sum_{i\neq j} \tau_{j,n}^*-c_n}{I-1}$, which leads to $t_{i,n}=0$. 	
	\begin{align}
	V_i(\bs{m}^*)&\geq V_i(\hat{\bs{m}}_i;\bs{m}_{-i}^*)=\!\! \max_{\bs{x}_i\in\tilde{\mathcal{X}}_i(\bs{m}_i;\bs{m}_{-i}^*)}\!\!U_i(\bs{x}_i)+\sum_{n\in\mathcal{N}_i}\frac{p_{n}^*c_n}{|\mathcal{I}_n|}.
	\end{align}

By Assumption \ref{Assum3}, agent $i$'s maximal payoff is
	\begin{align}
	V_i^{\rm Out}&\triangleq\max_{\bs{x}_i\in\mathcal{X}_i^{\rm Out}} U_i(\bs{x}_i).
	\end{align}	
	Note that $\mathcal{X}_i^{\rm Out}=\left\{\bs{x}_i: \bs{x}_i\in\mathcal{X}_i,~ h_{i,n}(\bs{x}_i)~\trianglelefteq_n~0,~\forall n\in\mathcal{N}\right\}=\tilde{\mathcal{X}}_i(\bs{m}_i;\bs{m}_{-i}^*)$. In addition,
	if $\trianglelefteq_n$ is $\leq$, then $p_n^*=\lambda_n^*\geq0$ and $c_n\geq0$ due to Assumption \ref{Assum5}; if $\trianglelefteq_n$ is $=$, then $c_n=0$ due to Assumption \ref{Assum5}. Therefore, we have that $\sum_{n\in\mathcal{N}}p_n^{*}c_n/|\mathcal{I}_n|\geq 0$. Thus, the GNE payoff for every agent $i$ always satisfies
	\begin{align}
	V_i(\bs{m}^*)\geq V_i^{\rm Out},
	\end{align}
	which demonstrates the individual rationality (E2).	


%

\subsection{Proof of Proposition \ref{T5}}\label{ProofP3}

Algorithm 1 performs in a similar fashion as the incremental subgradient method does in \cite{Incremental}. 
Specifically, from \cite{Incremental},  agents  update the dual variable $\bs\lambda$ to solve the dual problem in \eqref{DP}:
\begin{align}
\bs\lambda[k+1]=\tilde{\bs{p}}_{I}[k], \label{lambdaa}
\end{align}
with $\tilde{\bs{p}}_{i}[k]\in\mathbb{R}^N$ being the local dual variable obtained as
\begin{align}
&\tilde{\bs{p}}_{i}[k+1]\\
=&\begin{cases}
\tilde{\bs{p}}_{I}[k]-\alpha[k+1]d_{i,n}(\tilde{\bs{p}}_{I}[k]),&~~{\rm if}~~i=1,\\
\tilde{\bs{p}}_{i-1}[k+1]-\alpha[k+1]d_{i,n}(\tilde{\bs{p}}_{i-1}[k+1]),&~~{\rm if}~~i>1.
\end{cases}\nonumber
\end{align}
We define $\bs{z}_{i}(\bs\lambda)\in \partial g_i(\bs\lambda)$ as agent $i$'s local subgradient, given by
\begin{align}
z_{i,n}(\bs\lambda)=\begin{cases}
t_{i,n}^*-\frac{c_n}{|\mathcal{I}_n|},~~&{\rm if}~~i\in\mathcal{I}_n,\\
0,~~&{\rm otherwise},
\end{cases}
\end{align}
where
\begin{subequations}\label{Axq}
\begin{align}
(\bs{x}_i^*,\bs{t}_i^*)\in\arg&\max_{\substack{\bs{x}_i\in\mathcal{X}_i\\ \bs{t}_i}}\left\{U_i\left(\bs{x}_i\right)-\sum_{n\in\mathcal{N}_i}\bar{p}_{\omega(i-1,n),n}[k]t_{i,n}\right\}\\
&~{\rm s.t.}~~~h_{i,n}(\bs{x}_i)\trianglelefteq_n t_{i,n},~t_{i,n}\leq t_{i,n}^{\rm up},~\forall n\in\mathcal{N}_i.\label{Axq2}
\end{align}
\end{subequations}
The fact that $z_{i,n}=0$ if $i\notin\mathcal{I}_n$ indicates that
 \begin{align}\label{Axs}
 &\tilde{p}_{i,n}[k+1]\\
 =&\begin{cases}\tilde{p}_{\omega(i-1,n),n}[k]+\alpha[k+1]z_{i,n}(\tilde{\bs{p}}_{\omega(i-1,n)}[k]),\\
 ~~~~~~~~~~~~~~~~~~~~~~~~~~~~~~~~~~~~~~~~~{\rm if}~~i=\min_{j\in\mathcal{I}_n}j,\\
 \tilde{p}_{\omega(i-1,n),n}[k+1]+\alpha[k+1]z_{i,n}(\tilde{\bs{p}}_{\omega(i-1,n)}[k+1]),\\~~~~~~~~~~~~~~~~~~~~~~~~~~~~~~~~~~~~~~~~~~~~~~~~~~{\rm otherwise}.
 \end{cases}\nonumber
 \end{align}
Comparing \eqref{Axq}-\eqref{Axs} with \eqref{update1}-\eqref{update2}, we see that the above mentioned algorithm of the incremental subgradient method is equivalent to the DeNUM Algorithm.
 
Next, we prove that the subgradient $z_{i,n}(\bs\lambda)$ for every agent $i$ is bounded. Due to constraints in \eqref{update1} and \eqref{Axq2}, the subgradient $z_{i,n}(\bs\lambda)$ is bounded if $h_{i,n}(\bs{x}_i)> -\infty$ for all $\bs{x}_i\in\mathcal{X}_i$. Note that the compactness of ${\mathcal{X}}_i$ due to Assumption 1 ensures that $\{h_{i,n}\}$ are also bounded. This leads to the boundedness of the subgradients.

Therefore, to prove the convergence of Algorithm \ref{algo1}, it suffices to prove the convergence of the sequence $\{\bs\lambda[k]\}$ to the optimal dual variables of the dual problem $\min_{\bs\lambda}\sum_{i\in\mathcal{I}}g_i(\bs\lambda)$. We first adopt  the following lemma in \cite{Incremental}.

\begin{lemma}\label{L4}
	Let $\{\bs\lambda[k]\}$ be the sequence generated by \eqref{lambdaa}. We have that, for all $\bs{y}\in\mathbb{R}^{N}$ and $k\in\mathbb{N}$, 
	\begin{align}
	&||\bs{\lambda}[k+1]-\bs{y}||^2\nonumber\\
	&~~~~~~~~~\leq ||\bs{\lambda}[k]-\bs{y}||^2-2\alpha[k]\left(g(\bs{\lambda}[k])-g(\bs{y})\right)-\alpha[k]^2\hat{C}^2,\nonumber
	\end{align} 
	where $\hat{C}=\sum_{i\in\mathcal{I}}\sup_{\bs\lambda}{||g_i(\bs\lambda)||}$ because the subgradients $\bs{z}_i(\bs\lambda)$ are bounded.
\end{lemma}

We further adopt the following proposition in \cite{Incremental}:
\begin{proposition}\label{P6}
	With step size $\alpha[k]$ given by
	\begin{align}
	\alpha[k]>0,~~\sum_{k=0}^\infty\alpha[k]=\infty,~~~\sum_{k=0}^\infty\alpha[k]^2<\infty,
	\end{align}
	the sequence $\{\bs{\lambda}[k]\}$ converges to an optimal dual variable $\bs\lambda^o$.
\end{proposition}

Note that $||\tilde{\bs{p}}_{i}[k]-\bs{\lambda}[k]||\leq \alpha[k] \hat{C}$, which means that, for all $(i,n)\in\{(i,n): i\in\mathcal{I}_n\}$: 
\begin{align}
\lim_{k\rightarrow \infty}p_{i,n}[k]= \lambda_n^o,\label{DzoZ}
\end{align}	
where $\bs\lambda^o$ is the optimal dual variable to the dual problem in \eqref{DP}.
As we have discussed in Section \ref{ID}, when we substitute $\bs\lambda^o$ into the dual problem in \eqref{Indirect}, we also see that $\bs{x}_i[k]\rightarrow \bs{x}_i^o$ for every agent $i$. We complete the proof.

\subsection{Direct Decomposition and Proof of Proposition \ref{TT5}}\label{ProofP4}

In this part, we first present a direct decomposition structure of solving the NUM Problem. We then prove the Proposition \ref{TT5}.

\subsubsection{Direct Decomposition}\label{Dire}

We present the direct (dual) decomposition \cite{NUMTut} that serves as a distributed (pure) optimization method for solving the DeNUM Problem when agents are obedient.

To see this, we relax the constraint in \eqref{coupledconstraint} and \eqref{xi} and assign $\bs\lambda=\{\lambda_n\}_{n\in\mathcal{N}}$ to be the dual variables of it.  We can then formulate the corresponding Lagrangian of Problem in \eqref{NUM}, 
which can be further decomposed into $I$ locally solvable subproblems. We define the local dual function as follows:
That is, 
agent $i$'s local dual problem is: 
	\begin{align}\label{Direct}
	\tilde{g}_i(\bs\lambda)\triangleq&\max_{\bs{x}_i\in\mathcal{X}_i}~U_i(\bs{x}_i)-\sum_{n\in\mathcal{N}_i}\lambda_n \left(h_{i,n}(\bs{x}_i)-\frac{c_n}{|\mathcal{I}_n|}\right).
	\end{align}
At the higher layer, we obtain the optimal dual variable $\bs\lambda^o$ through solving a master (global) dual problem, given by
\begin{align}
\bs\lambda^o\in \arg\min_{\bs\lambda\in\Lambda}\sum_{i\in\mathcal{I}}\tilde{g}_i(\bs{\lambda}),\label{DDP}
\end{align}
where $\Lambda=\{\bs\lambda: \lambda_n\geq 0 ~{\rm for~each}~\trianglelefteq_n~{\rm being}~\leq\}$.

Substituting $\bs\lambda^o$ into \eqref{Direct}, we will have the optimal primary variables $\bs{x}_i^o$ for each agent $i$'s local problem. 
The above approach works only if agents are obedient.

\subsubsection{Proof of Proposition \ref{TT4}}
The DyDeNUM Mechanism together with updates in \eqref{Update1}-\eqref{Update3} performs in a similar fashion as the incremental subgradient method does in \cite{Incremental}. From \cite{Incremental},  agents  update the dual variable $\bs\lambda$ incrementally. Specifically, 
\begin{align}
\bs\lambda[k+1]=\tilde{\bs{p}}_{I}[k], \label{lambdaaa}
\end{align}
where $\tilde{\bs{p}}_{i}[k]\in\mathbb{R}^N$ is the local dual variable obtained
\begin{align}
&\tilde{\bs{p}}_{i}[k+1]\\
=&\begin{cases}
\tilde{\bs{p}}_{I}[k]-\alpha[k+1]z_{i,n}(\tilde{\bs{p}}_{I}[k]),&~~{\rm if}~~i=1,\\
\tilde{\bs{p}}_{i-1}[k+1]-\alpha[k+1]z_{i,n}(\tilde{\bs{p}}_{i-1}[k+1]),&~~{\rm if}~~i>1.
\end{cases}\nonumber
\end{align}
The vector function $\bs{z}_{i}(\bs\lambda)\in \partial \tilde{g}_i(\bs\lambda)$ is agent $i$'s local subgradient, given by
\begin{align}
z_{i,n}(\bs\lambda)=\begin{cases}h_{i,n}(\bs{d}_i[k])-\frac{c_n}{|\mathcal{I}_n|},&~{\rm if}~n\in\mathcal{N}_i,\\
0,&~{\rm otherwise}.
\end{cases}
\end{align}
The fact that $d_{i,n}=0$ if $i\in\mathcal{I}_n$ indicates that
\begin{align}
&\tilde{p}_{i,n}[k+1]\\
=&\begin{cases}\tilde{p}_{\omega(i-1,n),n}[k]+\alpha[k+1]z_{i,n}(\tilde{\bs{p}}_{\omega(i-1,n)}[k]),\\
~~~~~~~~~~~~~~~~~~~~~~~~~~~~~~~~~~~~~~~~~~{\rm if}~~i=\min_{j\in\mathcal{I}_n}j,\\
\tilde{p}_{\omega(i-1,n),n}[k+1]+\alpha[k+1]z_{i,n}(\tilde{\bs{p}}_{\omega(i-1,n)}[k+1]),\\~~~~~~~~~~~~~~~~~~~~~~~~~~~~~~~~~~~~~~~~~~~~~~~~~~~{\rm otherwise}.
\end{cases}\nonumber
\end{align}
Similar to the analysis in Section \ref{ProofP3} of this report, we can adopt Lemma \ref{L4} and Proposition \ref{P6} to prove that $\bs{d}_i[k]\rightarrow \bs{x}_i^o$ for each agent $i$.

\subsection{Proof of Theorem \ref{TT5}}\label{ProofT4}

Each agent $i$'s long-term average utility is
\begin{align}
&~~\lim_{K\rightarrow \infty}\frac{1}{K}\left[\sum_{k=1}^K U_i(\bs{d}_i[k])-\Pi_i[k]\right]\nonumber\\
&=\lim_{k\rightarrow \infty}[U_i(\bs{d}_i[k])-\Pi_{i}[k]]\nonumber\\
&= \lim_{k\rightarrow \infty}U_i(\bs{d}_i[k])-\Pi_{i}[0]\nonumber\\
&~~~~+\sum_{t=0}^\infty\sum_{j\neq i}\nabla_{\bs x_j} U_j(\bs{d}_j[k])^T(\bs{d}_{j}[k+1]-\bs{d}_{j}[k])\label{TT5-1}.
\end{align}
From updates in \eqref{Update1}-\eqref{Update3}, we have that,  for each $(i,n)\in\{(i,n): i\in\mathcal{I}_n\}$,
\begin{align}
|p_{i,n}[k+1]-p_{i,n}[k]|\leq \alpha[k]\left[\sum_{i\in\mathcal{I}_n} h_{i,n}[k]\right]
\leq \alpha[0]H_n,
\end{align}
where $H_n=\max_{\bs{x}_i\in\mathcal{X}_i}\sum_{i\in\mathcal{I}_n}h_{i,n}(\bs{x}_i)$  for all $n\in\mathcal{N}$. Due to the compactness of $\mathcal{X}_i$ by Assumption \ref{Assum2}, parameters $\{H_n\}_{n\in\mathcal{N}}$ always exist.
On the other hand, due to the strict concavity of the objective in \eqref{Update1}, we have that
$\bs{d}_i(\bs{p}_{\omega(i-1,n)})$ is continuous by the maximum theorem. The continuity indicates that $\lim_{\alpha[0]\rightarrow 0}||\bs{d}_i[k+1]-\bs{d}_i[k]||_2\rightarrow 0$. Therefore, \eqref{TT5-1} can be approximated as
\begin{align}
&\approxeq \lim_{k\rightarrow \infty}U_i(\bs{d}_i[k])-\Pi_{i}[0]+\sum_{j\neq i} \int_{0}^{\infty}\nabla_{\bs x_j} U_j(\cdot)^Td\bs{d}_{j}(k)\nonumber\\
&=\lim_{k\rightarrow \infty}\sum_{j\in\mathcal{I}}U_j(\bs{d}_j[k])-\Pi_{i}[0]+\sum_{j\neq i}  U_j(\bs{d}_j[0]).
\end{align}
By Proposition \ref{TT4}, the updates in \eqref{Update1}-\eqref{Update3} converge to the maximal network utility, i.e., $\bs\sum_{j\in\mathcal{I}}U_j(\bs{x}_j^o)$. This means that when all other agents are following the updates and reports according to \eqref{Update1}-\eqref{Update3}, there is no incentive for agent $i$ to deviate from following \eqref{Update1}-\eqref{Update3}. Hence, all agents following \eqref{Update1}-\eqref{Update3} is a Nash equilibrium.

\subsection{Distributed Computations of $\Pi_i[0]$ and Proof of Proposition \ref{P5}}\label{ProofP5}

In this subsection, we first design a distributed algorithm to compute the initial taxes $\Pi_i[0]$ so that each agent's final tax $\Pi_i[\infty]$ is a convergent VCG-type taxation. Then, we prove that such a VCG-type taxation can achieve the individual rationality.

\subsubsection{Distributed Computations of $\Pi_i[0]$}\label{Algo2}
	\begin{algorithm}[tb]\label{Algo3}		
	\SetAlgoLined
	 \caption{Distributive Computation of Initial Taxes}
	Initialize the iteration index  $k\leftarrow 0$ and $\Pi_i[0]$\;
	Each agent $j \in\mathcal{I}\backslash\{i\}$ randomly initializes $\{p_{\tilde{\omega}_i(j-1),n}[0]=C\}_{n\in\mathcal{N}_j}$ and the system designer chooses the stopping criterion $\epsilon$\;
	${\rm conv\_flag}\leftarrow 0$ $\#$ \textit{initialize the convergence flag}\; 
	\While{${\rm conv\_flag}= 0$ }{
		Set $k\leftarrow k+1$\; 
		\For{Each sub-iteration $j\in\mathcal{I}\backslash\{i\}$}{
			Each agent $j$ updates $(\bs{d}_j[k], {p}_{j,n}[k], \nabla u_j[k])$ by \eqref{Update11}-\eqref{Update33}\label{line777}\; 
	The system designer updates the $\Pi_i[0]$ as
	\begin{align}
		\Pi_{i}[0]=\Pi_{i}[0]+\sum_{j\in\mathcal{I}\slash\{i\}}\nabla u_j[k]^T(\bs{d}_{j}[k-1]-\bs{d}_{j}[k]).
\end{align}
			\eIf{$||\bs{p}_j[k]-\bs{p}_j[k-1]||_2\leq \epsilon||\bs{p}_{j}[k-1]||_2$}{
				${\rm conv\_count}\leftarrow {\rm conv\_count}+1$\;
			}{ ${\rm conv\_count}\leftarrow 0$\;}
			Each agent $i$ sends ${p}_{i,n}[k]$ to agent $\tilde{\omega}_i(i+1,n)$ for every $n\in\mathcal{N}_i$ and ${\rm conv\_count}$ to agent $i+1$\; 
		}
		\If{${\rm conv\_count}=I$}{
			Set ${\rm conv\_flag}\leftarrow 1$ and broadcasts it\;
		}		  	
	}
\end{algorithm}	
We define 
\begin{align}
&\tilde{\omega}_i(n,j+1)\\
\triangleq&\begin{cases}
\text{the $(\upsilon+1)$-th smallest index in $\mathcal{I}_n\backslash \{i\}$}, &{\rm if}~ \upsilon\neq|\mathcal{I}_n|-1, \\
\text{the smallest index in $\mathcal{I}_n\backslash \{i\}$},~&{\rm otherwise}.
\end{cases}\nonumber
\end{align}

    	Algorithm \ref{Algo3} shows the proposed iterative algorithm for all agents excluding agent $i$ to compute an appropriate $\Pi_i[0]$ for each agent $i$, with the key steps explained as follows. 
    	Each agent $j\neq i$ first  initializes her message $\bs{m}_i[0]\in\mathbb{R}^{2\times |\mathcal{N}_j|}$ (line \ref{l2}). Note that each agent initializes her price proposal $p_{j,n}[0]=C$ by the same constant in the DyDeNUM Mechanism.
    	The algorithm iteratively computes each agent's message and action until convergence (lines \ref{line4}-\ref{line188}). For each iteration, agents update their messages and actions in a Gauss-Seidel fashion (lines \ref{line6}-\ref{line14}). That is, we divide one iteration into $\mathcal{I}$ sub-iterations (line \ref{line6}). In each sub-iteration $i$, only agent $i$ updates her messages and actions,  whereas the other agents keep theirs fixed. 
    	
    	Specifically, in each sub-iteration $j$, each agent $j$ updates $(\bs{x}_j[k],\bs{\tau}_j[k],\bs{p}_j[k])$ (in line \ref{line777}) according to
    	\begin{align}
    	&\bs{d}_{j}[k]=\arg\max_{\bs{x}_j\in\mathcal{X}_j}\left[U_{j}(\bs{x}_j)-\sum_{n\in\mathcal{N}_j}\bar{p}_{\omega(j-1),n}[k] h_{j,n}(\bs{x}_j)\right],\label{Update11}~\\
    	&{p}_{j,n}[k]=\label{Update22}\\
    	&\begin{cases}\bar{p}_{\omega(j-1,n),n}[k]+\alpha[k]\left( h_{j,n}(\bs{d}_j[k])-\frac{c_n}{|\mathcal{I}_n|}\right),~{\rm if}~\trianglelefteq_n {\rm is~}=,\\\left[\bar{p}_{\omega(j-1,n),n}[k]+\alpha[k]\left( h_{j,n}(\bs{d}_j[k])-\frac{c_n}{|\mathcal{I}_n|}\right)\right]^+\!\!\!,~{\rm if}~\trianglelefteq_n {\rm is~}\leq,
    	\end{cases} \nonumber\\
    	&\nabla u_j[k]=\nabla_{\bs{x}_j} U_j(\bs{d}_j[k])\label{Update33},
    	\end{align}
    	where $\alpha[k]$ is a diminishing step size, given by $\alpha[k]=(1+h)/(k+h)$ for some non-negative constant $h$. Note that the updates in \eqref{Update11}-\eqref{Update33} are similar to the updates in \eqref{Update1}-\eqref{Update3}.
    	
    	Note that Algorithm \ref{Algo3} involves all agents other than $i$. Hence, since each agent  $i$ does not participate in the distributive computation of her own taxation, she cannot tamper with the algorithm to her advantage. Hence, we can assume that each agent will follow such an algorithm.
    	
Define $\hat{\bs{x}}^{(-i)}=\left\{\hat{\bs{x}}_j^{(-i)}\right\}_{j\in\mathcal{I}\backslash \{i\}}$ as the optimal solution of 
\begin{subequations}\label{OptSol}
\begin{align}
\hat{\bs{x}}^{(-i)}=\arg\max_{\bs{x}_{-i}}&~\sum_{j\neq i}~U_j(\bs{x}_j)\\
{\rm s.t. }&~\sum_{j\in\mathcal{I}\backslash\{i\}}h_{j,n}(\bs{x}_j)\trianglelefteq_n~c_n,~\forall n\in\mathcal{N}_i,\\
&~~\bs{x}_j\in\mathcal{X}_j,~\forall j\neq i, j\in\mathcal{I}.
\end{align}
\end{subequations}
The aggregate utility $\sum_{j\neq i}~U_j(\hat{\bs{x}}_j^{(-i)})$ is the maximal network utility when agent $i$ is absent. Similar to the argument in Section \ref{SecVII-B}, we can show that each agent $i$'s initial tax $\Pi_i[0]$ can be approximated in the following manner:
\begin{align}
\Pi_{i}[0]
&= \sum_{t=0}^\infty\sum_{j\neq i}\nabla_{\bs x_j} U_j(\bs{x}_j[k])^T(\bs{d}_{j}[k+1]-\bs{d}_{j}[k])\nonumber\\
&\approxeq \sum_{j\neq i} \int_{0}^{\infty}\nabla_{\bs x_j} U_j(\cdot)^Td\bs{d}_{j}(k)\nonumber\\
&=\sum_{j\neq i}  U_j(\hat{\bs{x}}^{(-i)}_j)-\sum_{j\neq i}  U_j(\bs{d}_j[0]).
\end{align}
Due to the same initializations of the price proposals, the value of $U_j(\bs{d}_j[0])$ is exactly the same as the value of it in \eqref{Approx}.
\subsubsection{Proof of Proposition \ref{P5}}

We next show that the initial tax computed by Algorithm \ref{Algo3}
for each agent $i$ leads to the individual rationality (E2).

The individual rationality of the DyDeNUM Mechanism is an immediate result of the constructed VCG taxation. Specifically, 
each agent's payoff at the equilibrium is given by
\begin{align}
J_i(\{\bs{m}^*[k]\}_{k\in\mathbb{N}})&=\sum_{i\in\mathcal{I}}U_i(\bs{d}_i[0])-\sum_{j\neq i} U_j(\hat{\bs{x}}_j^{(-i)})+\sum_{i\in\mathcal{I}}U_i(\bs{x}_i^o)\nonumber\\
&~~~-\sum_{i\in\mathcal{I}}U_i(\bs{d}_i[0]),\nonumber\\
&=\sum_{i\in\mathcal{I}}U_i(\bs{x}_i^o)-\sum_{j\neq i} U_j(\hat{\bs{x}}_j^{(-i)}).\label{Eq71}
\end{align}
As we have mentioned, due to the same initializations of the price proposals, the terms $\sum_{i\in\mathcal{I}}U_i(\bs{d}_i[0])$ are cancelled out.

By the definitions of $\mathcal{X}_i^{\rm Out}$ in \eqref{Out} and $\hat{\bs{x}}^{(-i)}$ in \eqref{OptSol}, we see that the action profile $\bs{x}=\{\bs{x}_i\}_{i\in\mathcal{I}}$ such that
\begin{align}
\bs{x}_i=\begin{cases}
\arg\max_{\bs{x}_i\in\mathcal{X}_i^{\rm Out}}U_i(\bs{x}_i),~{\rm for~some~}k,\\
\hat{\bs{x}}_j^{(-i)},~~\forall~j\neq k,
\end{cases}
\end{align}
%
 is a feasible solution to the DeNUM Problem in \eqref{NUM}. Hence, by the optimality of $\bs{x}^o$, we have 
\begin{align}
\sum_{i\in\mathcal{I}}U_i(\bs{x}_i^o)\geq \sum_{j\neq i} U_j(\hat{\bs{x}}_j^{(-i)})+\max_{\bs{x}_i\in\mathcal{X}_i^{\rm Out}}U_i(\bs{x}_i).
\end{align}
From \eqref{Eq71}, we have that $J_i(\{\bs{m}^*[k]\}_{k\in\mathbb{N}})\geq \max_{\bs{x}_i\in\mathcal{X}_i^{\rm Out}}U_i(\bs{x}_i)$. Hence, under Assumptions \ref{Assum4}-\ref{Assum5}, the DyDeNUM Mechanism achieves the individual rationality.

\begin{IEEEbiography}[]{Meng Zhang} (S'15) is a Ph.D.
	student in the Department of Information Engineering  at the Chinese
	University of Hong Kong since 2015. He was a visiting student research collaborator with the Department of Electrical Engineering at Princeton University, from 2018 to 2019.
	His research interests include network
	economics, with emphasis on pricing and mechanism design for networked systems.
\end{IEEEbiography}

\begin{IEEEbiography}[]{Jianwei Huang} (F'16) is a Presidential Chair Professor and the Associate Dean of the School of Science and Engineering, The Chinese University of Hong Kong, Shenzhen. He is also the Associate Director of Shenzhen Institute of Artificial Intelligence and Robotics for Society (AIRS), and a Professor in the Department of Information Engineering, The Chinese University of Hong Kong. He received the Ph.D. degree from Northwestern University in 2005, and worked as a Postdoc Research Associate at Princeton University during 2005-2007. He has been an IEEE Fellow, a Distinguished Lecturer of IEEE Communications Society, and a Clarivate Analytics Highly Cited Researcher in Computer Science. He is the co-author of 9 Best Paper Awards, including IEEE Marconi Prize Paper Award in Wireless Communications in 2011. He has co-authored six books, including the textbook on "Wireless Network Pricing." He received the CUHK Young Researcher Award in 2014 and IEEE ComSoc Asia-Pacific Outstanding Young Researcher Award in 2009. He has served as an Associate Editor of IEEE Transactions on Mobile Computing, IEEE/ACM Transactions on Networking, IEEE Transactions on Network Science and Engineering, IEEE Transactions on Wireless Communications, IEEE Journal on Selected Areas in Communications - Cognitive Radio Series, and IEEE Transactions on Cognitive Communications and Networking. He has served as an Editor of Wiley Information and Communication Technology Series, Springer Encyclopedia of Wireless Networks, and Springer Handbook of Cognitive Radio. He has served as the Chair of IEEE ComSoc Cognitive Network Technical Committee and Multimedia Communications Technical Committee. He is the Associate Editor-in-Chief of IEEE Open Journal of Communications Society. He is the recipient of IEEE ComSoc Multimedia Communications Technical Committee Distinguished Service Award in 2015 and IEEE GLOBECOM Outstanding Service Award in 2010. More detailed information can be found at http://jianwei.ie.cuhk.edu.hk/.
\end{IEEEbiography}

\end{document}